\documentclass[lineno]{jfm}

\usepackage{graphicx}
\usepackage{makecell}
\usepackage{newtxtext}
\usepackage{newtxmath}
\usepackage[dvipsnames]{xcolor}
\usepackage{natbib}
\usepackage{hyperref}
\usepackage{comment}
\hypersetup{
    colorlinks = true,
    linkcolor  = blue,
    urlcolor   = blue,
    citecolor  = blue,
}

\newcommand{\RomanNumeralCaps}[1]
\linenumbers

\title{Rough surfaces in under-explored surface morphology space and their implications on roughness modelling}

\author{Shyam S. Nair\aff{1}
  \corresp{\email{shyam.nair@psu.edu}},
   Vishal A. Wadhai\aff{1}, Robert F. Kunz\aff{1} \and Xiang I. A. Yang\aff{1}
  }

\affiliation{\aff{1}Mechanical Engineering, Penn State University, State College, PA 16802, USA}

\begin{document}
\maketitle

\begin{abstract}
We report direct numerical simulation (DNS) results of the rough-wall channel, focusing on roughness with high $k_{rms}/k_a$ statistics but small to negative $Sk$ statistics, and we study the implications of this new dataset on rough-wall modelling. 
Here, $k_{rms}$ is the root-mean-square, $k_a$ is the first order moment of roughness height, and $Sk$ is the skewness. 
The effects of packing density, skewness and arrangement of roughness elements on mean streamwise velocity, equivalent roughness height ($z_0$) and Reynolds and dispersive stresses have been studied.
We demonstrate that two-point correlation lengths of roughness height statistics play an important role in characterizing rough surfaces with identical moments of roughness height but different arrangements of roughness elements.  
Analysis of the present as well as historical data suggests that the task of rough-wall modelling is to identify geometric parameters that distinguish the rough surfaces within the calibration dataset.
We demonstrate a novel feature selection procedure to determine these parameters. 
Further, since there is not a finite set of roughness statistics that distinguish between all rough surfaces, we argue that obtaining a universal rough-wall model for making equivalent sand-grain roughness ($k_s$) predictions would be challenging, and that each rough-wall model would have its applicable range. 
This motivates the development of group-based rough-wall models. The applicability of multi-variate polynomial regression and feedforward neural networks for building such group-based rough-wall models using the selected features has been shown. 
\end{abstract}



\section{Introduction}
\label{sec:headings}

Turbulent flow over rough walls has piqued the interest of researchers and engineers for decades. 
From the earliest seminal works by \cite{nikuradse1933stromungsgesetze}, 
\cite{colebrook1939correspondence} and the consolidated data interpretation by \cite{moody1944friction}, the field of rough-wall bounded turbulence has continued to evolve, as evident in subsequent works \citep{raupach1991rough, jimenez2004turbulent, flack2010review}. 
In the recent review by \cite{Chung2021PredictingSurfaces}, the authors emphasize the importance of Townsend's outer-layer similarity hypothesis \citep{townsend1976structure}.
This hypothesis assumes that, under fully rough and sufficiently high boundary layer thickness to roughness height ratio ($\delta /k$) conditions, the outer layer of a rough wall boundary layer behaves similarly to that of a smooth wall. 
A logarithmic layer can therefore be expected between the roughness sublayer and the outer layer. 
This logarithmic layer, aside from a downward shift $\Delta U^+$ compared to the smooth wall log law, remains unaffected by surface roughness.
This downward shift $\Delta U^+$ constitutes the roughness function, enabling frictional drag comparisons across different rough surfaces. 
Besides $\Delta U^+$, which characterizes the hydraulic properties, a rough wall is also characterized by the statistical moments of roughness height and the distribution of its topographical features. 
Establishing a functional mapping between the roughness function and the rough wall topography is of significant practical utility and remains a challenging problem, given the variations in surface features and the costs of rough wall boundary layer experiments \citep{schultz2007rough} and scale-resolving computational studies \citep{yang2021grid,choi2012grid}. 

Several studies have aimed to develop roughness correlations for various surfaces. 
\cite{flack2010review} proposed correlations utilizing skewness ($Sk$) and root-mean-square roughness height ($k_{rms}$) to characterize equivalent sand-grain roughness ($k_s$) for irregular three-dimensional roughness.
\cite{yuan2014estimation} examined slope-based and moment-based correlations to study critical effective slope ($ES$) associated with waviness regimes for realistic roughness from hydraulic turbine blades.
\cite{Forooghi2017TowardCorrelation} found that functional relations dependent on $Sk$ and $ES$, which represent the shape and slope of the rough surface respectively, produced satisfactory $k_s$ predictions for randomly distributed roughness elements of random size and prescribed shape.
In another study, \cite{flack2020skin} performed experiments on random rough surfaces by systematically varying $k_{rms}$ and $Sk$. 
They found predictive correlations of the form $k_s = Ak_{rms}(1+Sk)^B$ with $A$ and $B$ being constants. 
The authors also adapted the functional form according to groups of surfaces as being either positively, negatively, or zero-skewed for more accurate correlations. 
On the other hand, correlations can also take a data-driven form as developed by \cite{AghaeiJouybari2021Data-drivenFlows}, where deep neural networks (DNN) incorporate information from diverse rough-wall geometries and their corresponding statistical moments.
Irrespective of the models used, no single model has been able to generalize well across all rough surfaces 
\citep{yang2023search}.

The difficulty is largely due to the complexity of rough surfaces and that each rough surface seems to have its unique behaviours.
This calls for the categorization of rough surfaces.
A possible categorization puts rough surfaces into regular roughness and irregular roughness. 
Regular surfaces have the same elements repeated in a predefined periodic arrangement, unlike irregular roughness where the features are random in shape and/or distribution.
Categories based on the shape of the roughness features break down as being cubes \citep{castro2006turbulence}, truncated cones \citep{womack2022turbulent}, packed spheres \citep{schultz2005outer}, grit-blasted \citep{flack2016skin, thakkar_busse_sandham_2018} and others. 
Rough surfaces may also be categorized by distribution, such as Gaussian \citep{flack2020skin, ma2021direct, altland2022flow}, random (randomly distributed regular roughness elements with Gaussian height distributions)  \citep{Forooghi2017TowardCorrelation, forooghi2018direct}, pseudo-random  \citep{yang2022direct}, multiscale \citep{yang2017modelling, medjnoun2021turbulent} and so on.
The distinction between roughness types could also stem from its flow physics. 
For instance, the k-type and d-type roughness \citep{jimenez2004turbulent} exhibit different behaviours and correspond to sparse and densely packed roughness elements respectively.

The more recent literature seems to favor more precise categorizations based on roughness' statistics.
\cite{Chung2021PredictingSurfaces} identified several surface properties to be hydrodynamically important. 
These include measures of roughness height (such as root-mean-square $k_{rms}$, average $k_a$, maximum peak-to-trough $k_t$), effective slope ($ES$), frontal solidity ($\lambda_f$), planar packing density ($\lambda_p$), skewness ($Sk$), solid volume fraction ($\phi$) among others. 
A number of existing studies focus on the effects of variations in these individual parameters.
\cite{placidi2015effects} studied the effects of varying $\lambda_p$ and $\lambda_f$ with ``LEGO'' brick-type roughness in the fully rough regime. 
Unlike cubical roughness, they found roughness length to monotonically decrease with increasing $\lambda_p$, suggesting that the same geometrical parameter may behave differently based on roughness type. 
\cite{thakkar2017surface} found that the streamwise correlation length plays a role in determining the roughness function in addition to $k_{rms}$, $Sk$, and $\lambda_f$ for irregular rough surfaces including samples that are cast, composite, hand filed, grit blasted, ground, spark-eroded and from ship propellers.
The effect of surface anisotropy was systematically investigated by \cite{busse2020influence} for irregular surfaces where another roughness parameter, the surface anisotropy ratio (SAR), defined as the ratio of the streamwise and spanwise correlation lengths, was varied and found to strongly influence the mean flow.
Even spanwise parameters such as the spanwise spacing \citep{vanderwel_ganapathisubramani_2015} and spanwise effective slope ($ES_y$) \citep{jelly2022impact} are found to significantly affect mean flow statistics. 
It appears that whenever one varies a roughness parameter, that roughness parameter stands out and plays an important role in determining the equivalent sandgrain roughness height.

In anticipation of the discussion in the following sections, here we review the previous studies of cubical roughness.
Due to its simplicity, cubical roughness is one of the most extensively studied types of surface roughness.
Early experimental investigations on cube patterns \citep{o1964some} studied the effect of $\lambda_p$ on equivalent roughness size ($k_s/k$), establishing that the resistance to flow reaches a maximum at an intermediate $\lambda_p$, around 0.2, and then tends towards smooth wall behaviour at increasing cube densities.   
Thanks to the rich physics cube arrays can represent, several studies have enriched the cubical roughness literature with details about its roughness sublayer \citep{castro2006turbulence}, aerodynamic characteristics \citep{cheng2007flow} and associated flow structures \citep{volino2011turbulence}. 
Computational studies implementing Direct Numerical Simulations (DNS) have further prompted discussions on the mean velocity profile, equivalent roughness height ($z_0$), and zero-plane displacement height ($d$).
These include works by \cite{leonardi2010channel} and \cite{lee2011direct} which focused on turbulence statistics and coherent structures respectively.  
The utility involved in studying cubical roughness arising from its ability to generate surface parametrizations and the relevance to urban canopy studies have also led to analytical roughness models by \cite{yang2016large} and \cite{yang2016exponential}.

\begin{figure}
\centerline{\includegraphics[width=0.8\linewidth]{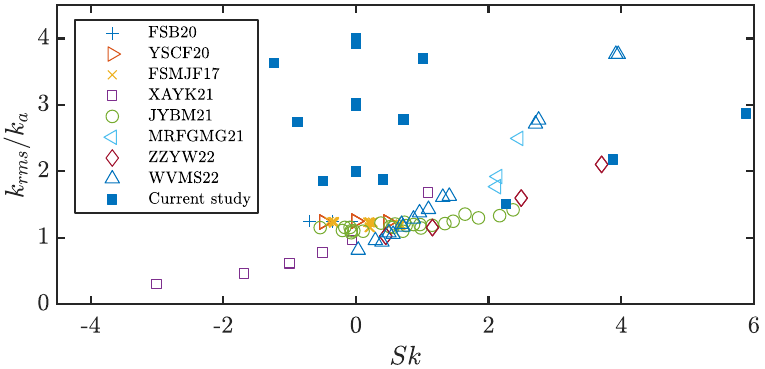}}
  \caption{Parameter space describing the rough surfaces in the current study and certain existing literature: \cite{flack2020skin}, \cite{yang2022direct}, \cite{Forooghi2017TowardCorrelation}, \cite{xu2021flow}, \cite{AghaeiJouybari2021Data-drivenFlows}, \cite{medjnoun2021turbulent}, \cite{Zhang2022EvidenceFlows} \& \cite{womack2022turbulent}. Each dataset is denoted by the initials of the authors’ last names.}
\label{fig:para}
\end{figure}

Based on the existing literature, it can be argued that surface features $Sk$ and $k_{rms}$ are two of the most important parameters and that a parameter space involving these features would be significant to look at. 
Figure \ref{fig:para} highlights the context of this study with respect to the existing literature in the $Sk-k_{rms}/k_a$ space.
It can be seen that most studies have focused on surfaces with either low $Sk$ and low $k_{rms}/k_a$ or high $Sk$ and high $k_{rms}/k_a$. 
In this study, we will expand the investigated parameter space by designing unique cubical surfaces that fall in a region of low $Sk$ and high $k_{rms}/k_a$.
This data set will allow us to study the hydrodynamic properties of unconventional roughness and test the efficacy of the existing rough-wall modelling approaches.
We shall see that the valleys, or the pits, do not contribute significantly to any of the first and second-order statistics reported here.
We shall also see that the accuracy of a rough-wall model largely depends on whether its input space distinguishes the rough walls under consideration.

The rest of the paper is organized as follows.
We present the details of the computational setup in \S\ref{sec:2} along with the geometry of the rough surfaces. 
The DNS results, including the instantaneous flow-field and the mean flow statistics, are presented in \S\ref{sec:3}.
We shall see that although the rough surfaces presented here are new, the resulting flow behavior aligns with trends found in existing literature.
In \S\ref{sec:4}, we discuss the implications of the DNS results on roughness modelling.
Finally, conclusions are given in \S\ref{sec:5}.

\section{Computational details}\label{sec:2}
\subsection{Case setup}

\begin{figure}
  \centerline{\includegraphics[scale=0.5]{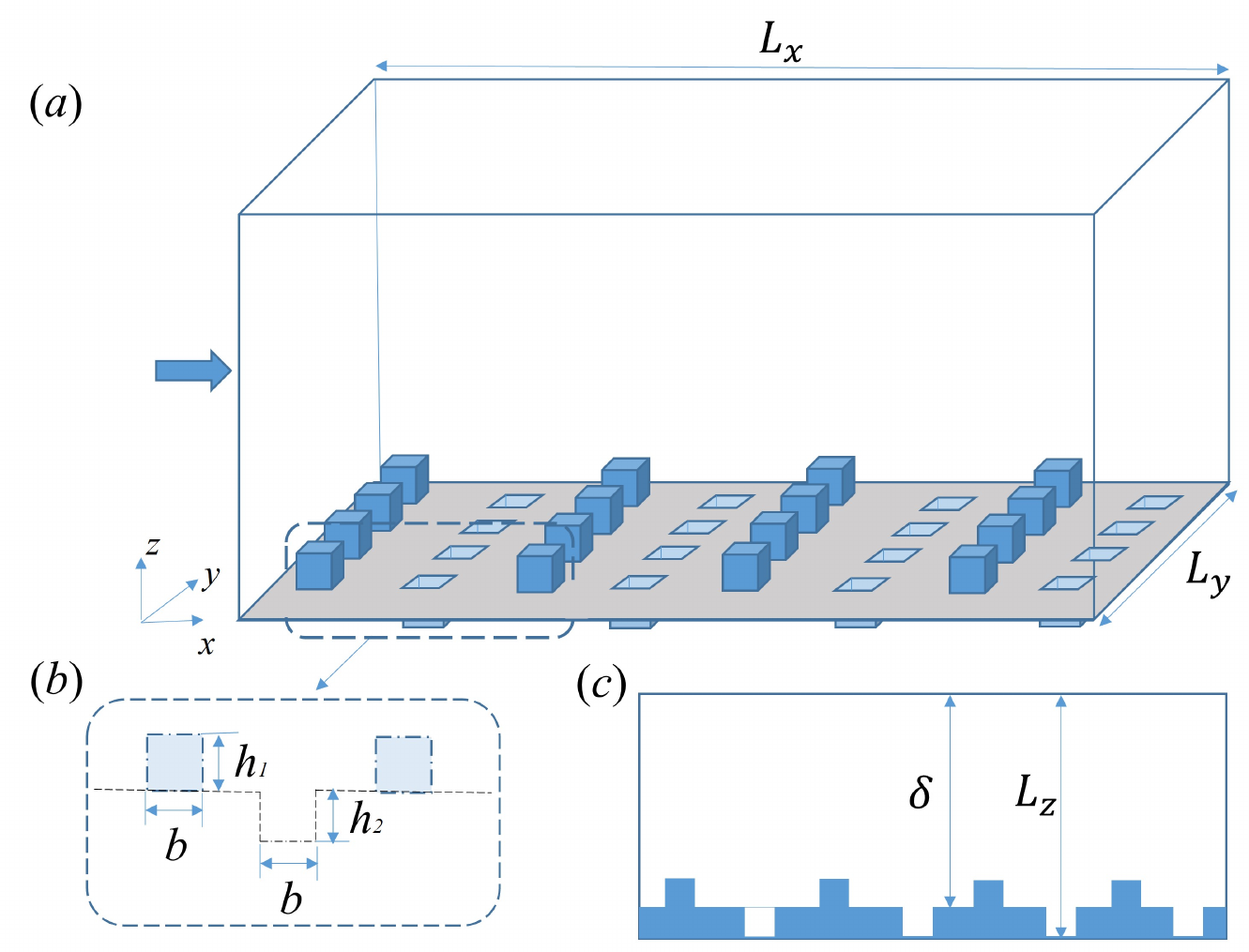}}
  \caption{A sketch illustrating a periodic half-channel flow setup over cubelike peaks and valleys. Here $h_1$ denotes peak height, $h_2$ denotes, valley depth and $b$ denotes the width of the cubelike element. The boundary layer thickness $\delta$ here is at least 6 times the maximum of $h_1$, and $h_2$.}
\label{fig:sketch}
\end{figure}

Figure \ref{fig:sketch} depicts a half-channel setup containing a wall with cube-like roughness elements. 
This is a representative domain over which flow is computed for the DNS runs. 
The bottom surface is a rough wall comprising pits (or valleys) and protrusions (or peaks) in the form of cube-like elements.
The $x$, $y$ and $z$ represent streamwise, spanwise and wall-normal directions.
Periodic conditions are applied to the lateral boundaries. A stress-free condition with no penetration (${\partial u}/{\partial z} = {\partial v}/{\partial z} = w = 0$) is imposed on the upper boundary.
A streamwise pressure gradient acts as the forcing that drives the flow. 
A friction Reynolds number of $Re_\tau \approx 400$  is used for all cases in this study, where $Re_\tau$ is defined as 
\begin{equation} \label{eq:1}
    R e_\tau=\frac{u_\tau (L_z-h_2)}{\nu}.
\end{equation}
Note that the wall-normal length $L_z$ used in \eqref{eq:1} includes the depth of the negatively skewed features or valleys $h_2$.
This depth $h_2$ is subtracted from $L_z$ to make a good estimate of the half-channel height.
The friction velocity $u_\tau$ is obtained using \eqref{eq:2} 
\begin{equation} \label{eq:2}
u_\tau= \sqrt{\left|\frac{\text{d} \langle \overline{p}\rangle }{\text{d} x}\right|\frac{V_f}{\rho A_p}} \approx \sqrt{\left|\frac{\text{d} \langle \overline{p}\rangle }{\text{d} x}\right|\frac{(L_z-h_2)}{\rho}} 
\end{equation}
where d$\langle \overline{p}\rangle$/d$x$ refers to the mean streamwise pressure gradient and $\rho$, $V_f$, $A_p$ refer to the fluid density, half-channel fluid volume and wall-parallel area respectively. 
Here $\langle .\rangle$ denotes the streamwise and spanwise averaging (or double-averaging) operation and $\overline{(.)}$ denotes time averaging. 
Note that $V_f$ includes the fluid volume within valleys and the approximation sign in \eqref{eq:2} is used because the peak-height is not always equal to the valley-depth, causing minor differences in $V_f$ in some cases.

The size of the computational domain is determined such that $L_x>6L_z$, $L_y>3L_z$.
\cite{lozano2014effect} have shown that such a domain would be sufficiently large to ensure the accuracy of first and second-order statistics for plane channel flow.
Note that $L_x$ and $L_y$ represent the streamwise and spanwise domain lengths respectively.
For rough-wall flows, this domain size should be a conservative estimate since DNS studies in \cite{coceal2006mean} and \cite{leonardi2010channel} produce good mean flow statistics with smaller domains of $L_x \times L_y \times L_z = 4h \times 4h \times 4h$ and $8h \times  6h \times 8h$ respectively ($h$ being the height of the cube). 
Other studies, such as \cite{chung2015fast}, have shown that even further reduction in computational domain is possible in the spanwise direction and such minimal-span channels have been shown to capture accurate mean drag characteristics (at least for sinusoidal roughness) when $L_y > k/0.4$ and $L_y^+>100$. 
Furthermore, since the roughness is regular, $L_x$ and $L_y$ are also integral multiples of the length of the repeating tile to ensure a periodic domain. 
Here, a repeating tile represents the smallest unit which when repeated in the streamwise and spanwise directions produces the entire surface.

A uniform grid has been utilized with grid resolution such that $\Delta_x^+ = \Delta_y^+<5$ and $\Delta_z^+<3$, where these represent streamwise, spanwise and wall-normal grid spacings respectively normalized by the viscous length scale ($\delta_v = \nu/u_\tau$).
It is important to note that while the $\Delta_z^+$ may appear to be large for DNS, this resolution is based on friction velocity $u_{\tau}$ obtained from \eqref{eq:2}. 
The $\delta_v$ computed based on this $u_{\tau}$ includes the form drag component. 
If $\delta_v$ is computed from the local viscous stress at the mean surface comprising the bottom wall, the grid resolution is at most 0.66 plus units.
Note that the local viscous stress $\mu \partial{u}/\partial{n}$ here does not pertain to the cube surfaces, where the corresponding value would spike at the leading edge due to the strong shear rate.
A similar grid resolution of $\Delta_x^+ = \Delta_y^+<4.5$ and $\Delta_z^+\approx 2.5$ was used by \cite{Zhang2022EvidenceFlows} for deep canopy flows.
The grid spacing is also comparable to that in the previous studies of rough wall DNS, where \cite{coceal2006mean} used $\Delta ^+_{x,y,z}$ in the range 7.8 to 15.6, and \cite{leonardi2010channel} used $\Delta ^+_{x,y} = 19$.
For further reference, DNS studies by \cite{Forooghi2017TowardCorrelation} have found $\Delta ^+_{x,y,z_{max}} = 3.5$ to be more than sufficient.
\cite{yuan2014roughness} utilize $\Delta ^+_{x} = 12$ and $\Delta ^+_{y} = 6$ whereas \cite{AghaeiJouybari2021Data-drivenFlows} employ $\Delta ^+_{x} = 7.5 $ and $\Delta ^+_{y} = 6.3$ for obtaining accurate flow statistics in their rough wall channels.

All simulations are performed using LESGO: A parallel pseudo-spectral large-eddy simulation code (\url{https://lesgo-jhu.github.io/lesgo}), a solver of the filtered Navier-Stokes with pseudo-spectral discretization in the streamwise and spanwise directions and a second-order finite difference in the wall-normal direction. 
LESGO and modified versions of the code have been extensively used for studies involving channel flows including \cite{bou2005scale}, \cite{anderson2011dynamic}, \cite{abkar2012new}, \cite{zhu2017parametric}, \cite{yang2019drag} and several others. 
Immersed boundary conditions \citep{anderson2013immersed} are used at the solid boundaries comprising the rough wall.
A Courant–Friedrichs–Lewy (CFL = $u\Delta t/\Delta x$) number of 0.2 is employed to automatically adjust the time-step, which is advanced via the second-order Adams–Bashforth scheme.

\subsection{Roughness generation}

\begin{figure}
  \centerline{\includegraphics[width=0.8\linewidth]{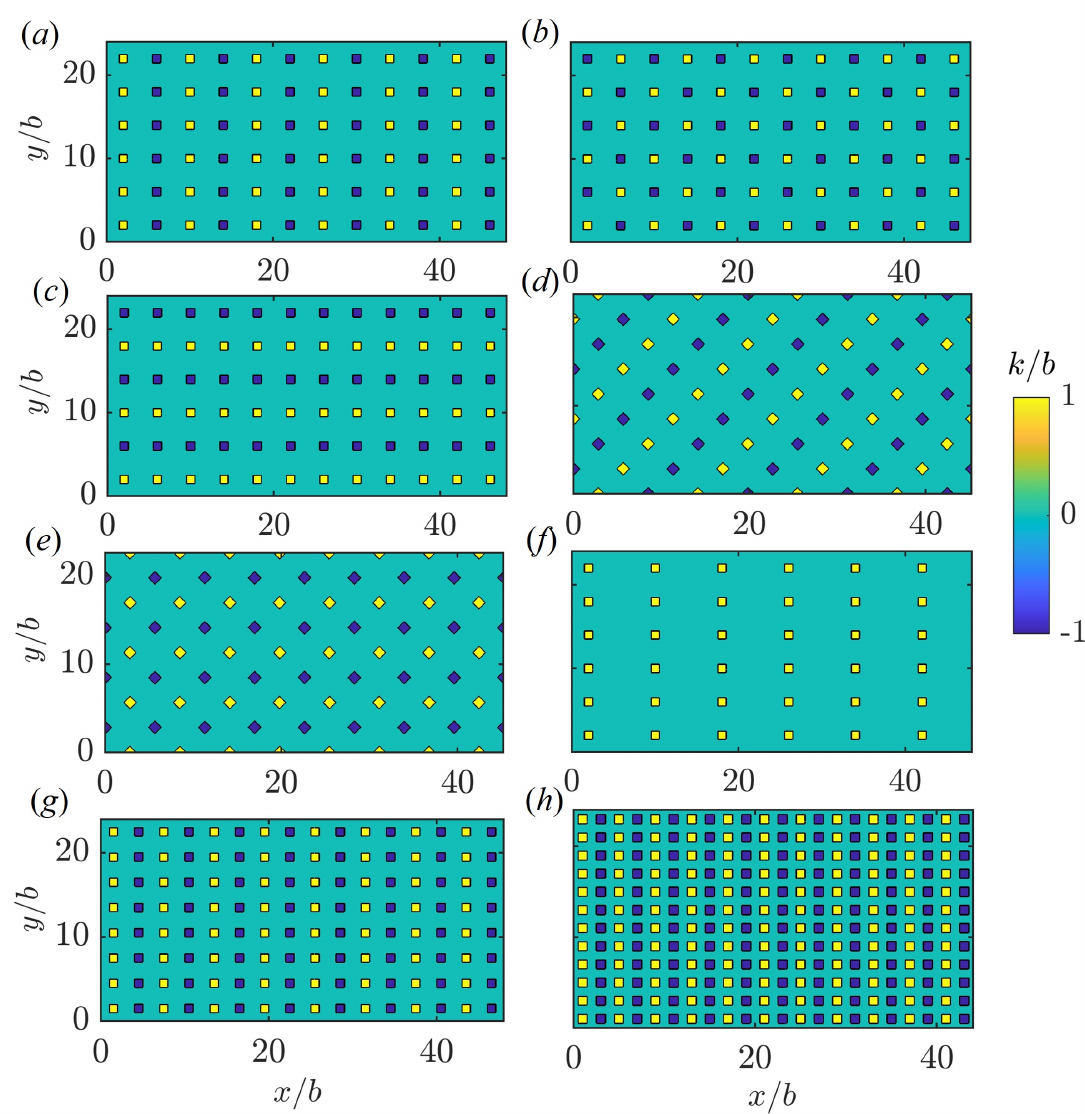}}
  \caption{Top view of various rough-surface types considered for the DNS study: (\textit{a}) AYL1H1, (\textit{b}) SL1H1, (\textit{c}) AXL1H1, (\textit{d}) AXYL1H1, (\textit{e}) SXYL1H1, (\textit{f}) NVL1H1, (\textit{g}) AYL2H1 and (\textit{h}) AYL3H1. $k$ is the elevation of the surface. The peaks have positive (depicted as yellow) and the valleys have negative $k$ (depicted as 
  blue).}
\label{fig:typical_cases}
\end{figure}

Roughness elements are distributed in various arrangements to generate different rough surfaces. 
We will vary one statistic at a time but repeat the process and vary many roughness statistics.
Figures \ref{fig:typical_cases}(\textit{a}), (\textit{c}) and (\textit{d}) depict these elements aligned in the spanwise (AY), streamwise (AX) and $45^\circ$ (AXY) directions respectively. 
Their staggered equivalents (S and SXY) are shown in figures \ref{fig:typical_cases}(\textit{b}) and (\textit{e}).
Figure \ref{fig:typical_cases}(\textit{f}) shows another surface (NV) where the negatively skewed features(valleys) are removed.
The rough surfaces also include variations in $\lambda_p$, as indicated by figures \ref{fig:typical_cases}(\textit{g} and \textit{h}), peak height-to-width ratio ($h_1/b$), and valley depth-to-width ratio ($h_2/b$). 
Table \ref{tab:Cases_nomenclature} provides the naming convention used and table \ref{tab:Cases_setup} lists all 36 different surfaces considered for the DNS study.
For each case, the nomenclature is as follows: [Arrangement][Packing density][Roughness height], where arrangement could be AX, AY, AXY, S, SXY, NV, packing density could be L1, L2, L3, and roughness height could be H1, H2, H3. 
For example, the L1, L2 and L3 in AYL1H1, AYL2H1 and AYL3H1 correspond to $\lambda_p$ magnitudes of 6.25, 11.1 and 25 percent respectively.
Similarly, the H1, H2 and H3 in AYL1H1, AYL1H2 and AYL1H3 denote variations in the height of the peaks  as $h_1 =  b, 1.2b$ and $0.8b$.
Note that, in this context, the conditions $h_1/h_2 = 1$, $h_1/h_2 > 1$ and $h_1/h_2 < 1$, correspond to zero-skewed, positively-skewed and negatively-skewed surfaces.
The separation length between the roughness elements is listed as $4b$, $3b$ and $2b$ for L1, L2 and L3 configurations with $b/L_z$ assuming values of 0.1375, 0.1111 and 0.1429 for H1, H2 and H3 cases respectively.

The definitions for a few roughness statistics $k_{rms}$, $k_a$, $k_t$, $Sk$, $ES$, $Ku$ and $\lambda_p$ (which are further discussed in table \ref{tab:DNS results}) for the rough surfaces thus generated are listed as follows:

\begin{equation} \label{eq:7}
    \overline{k}=\frac{1}{L_x L_y}\int_{L_x}\int_{L_y} k(x,y)\partial{x}\partial{y}
\end{equation}

\begin{equation} \label{eq:8}
    k_a=\frac{1}{L_x L_y}\int_{L_x}\int_{L_y} |k(x,y)-\overline{k}|\partial{x}\partial{y}
\end{equation}

\begin{equation} \label{eq:9}
    k_{rms}=\sqrt{\frac{1}{L_x L_y}\int_{L_x}\int_{L_y}|k(x,y) - \overline{k}|^2\partial{x}\partial{y}}
\end{equation}

\begin{equation} \label{eq:9-2}
    k_{t}=\text{max}(k(x, y)) - \text{min}(k(x,y))
\end{equation}

\begin{equation} \label{eq:10}
    Sk=\frac{1}{L_x L_y}\int_{L_x}\int_{L_y} (k(x,y)-\overline{k})^3\partial{x}\partial{y}/k_{rms}^3
\end{equation}

\begin{equation} \label{eq:11}
    Ku=\frac{1}{L_x L_y}\int_{L_x}\int_{L_y} |k(x,y)-\overline{k}|^4\partial{x}\partial{y}/k_{rms}^4
\end{equation}

\begin{equation} \label{eq:12}
    ES=\frac{1}{L_x}\int_{L_x} \Big|\frac{\partial k}{\partial x}\Big|\partial x
\end{equation}

\begin{equation} \label{eq:12.1}
    \lambda_p=\frac{A_p}{L_x L_y}
\end{equation}

\begin{table}
  \begin{center}
\def~{\hphantom{0}}
  \begin{tabular}{lll}
      Arrangement  & Packing density &   Roughness height\\[3pt]
       AX: streamwise-aligned   & L1: 6.25\%(AX, AY, AXY, S, SXY) & H1: $h_1 = b$ \\
       AY: spanwise-aligned   &  L1: 3.13\%(NV) & H2: $h_1 =1.2b$ \\
       AXY: $45^\circ$-aligned   & L2: 11.1\% (AX, AY, AXY, S, SXY) & H3: $h_1 =0.8b$ \\
       S: staggered   & L2: 5.55\% (NV) &  \\
       SXY: $45^\circ$-staggered   & L3: 25\% (AX, AY, AXY, S, SXY)    &  \\
       NV: no-valley & L3: 12.5\% (NV)&  \\
 \end{tabular}
 
  \caption{Nomenclature for the 36 configurations of the parameter space studied.}
  \label{tab:Cases_nomenclature}
  \end{center}
\end{table}

\begin{table}
  \begin{center}
\def~{\hphantom{0}}
  \begin{tabular}{lcccccc}
      Case  & $\lambda_p$   &   $h_1/b$ & $h_2/b$ & $N_x \times N_y \times N_z$ & $L_x/b \times L_y/b \times L_z/b$ & $N$\\[3pt]
       AYL1H1   & 6.25 & ~~1.0~ & ~~1.0~ & $672 \times 336 \times 160$ & $48.0 \times 24.0 \times 7.3$ & $36 + 36$\\
       AYL2H1   & 11.1 & ~~1.0~ & ~~1.0~ & $672 \times 336 \times 160$ & $48.0 \times 24.0 \times 7.3$ & $64 + 64$\\
       AYL3H1   & 25.0 & ~~1.0~ & ~~1.0~ & $616 \times 336 \times 160$ & $44.0 \times 24.0 \times 7.3$ & $132 + 132$\\
       AYL1H2   & 6.25 & ~~1.2~ & ~~1.0~ & $672 \times 384 \times 180$ & $56.0 \times 32.0 \times 9.0$ & $56 + 56$\\
       AYL2H2   & 11.1 & ~~1.2~ & ~~1.0~ & $648 \times 360 \times 180$ & $54.0 \times 30.0 \times 9.0$ & $90 + 90$\\
       AYL3H2   & 25.0 & ~~1.2~ & ~~1.0~ & $672 \times 336 \times 180$ & $56.0 \times 28.0 \times 9.0$ & $196 + 196$\\
       AYL1H3   & 6.25 & ~~0.8~ & ~~1.0~ & $672 \times 336 \times 175$ & $48.0 \times 24.0 \times 7.0$ & $36 + 36$\\
       AYL2H3   & 11.1 & ~~0.8~ & ~~1.0~ & $588 \times 336 \times 175$ & $42.0 \times 24.0 \times 7.0$ & $56 + 56$\\
       AYL3H3   & 25.0 & ~~0.8~ & ~~1.0~ & $616 \times 336 \times 175$ & $44.0 \times 24.0 \times 7.0$ & $132 + 132$\\
       SL1H1   & 6.25 & ~~1.0~ & ~~1.0~ & $672 \times 336 \times 160$ & $48.0 \times 24.0 \times 7.3$ & $36 + 36$ \\
       SL2H1   & 11.1 & ~~1.0~ & ~~1.0~ & $672 \times 336 \times 160$ & $48.0 \times 24.0 \times 7.3$ & $64 + 64$\\
       SL3H1   & 25.0 & ~~1.0~ & ~~1.0~ & $616 \times 336 \times 160$ & $44.0 \times 24.0 \times 7.3$ & $132 + 132$\\
       SL1H2   & 6.25 & ~~1.2~ & ~~1.0~ & $672 \times 384 \times 180$ & $56.0 \times 32.0 \times 9.0$ & $56 + 56$\\
       SL2H2   & 11.1 & ~~1.2~ & ~~1.0~ & $648 \times 360 \times 180$ & $54.0 \times 30.0 \times 9.0$ & $90 + 90$\\
       SL3H2   & 25.0 & ~~1.2~ & ~~1.0~ & $672 \times 336 \times 180$ & $56.0 \times 28.0 \times 9.0$ & $196 + 196$\\
       SL1H3   & 6.25 & ~~0.8~ & ~~1.0~ & $672 \times 336 \times 175$ & $48.0 \times 24.0 \times 7.0$ & $36 + 36$\\
       SL2H3   & 11.1 & ~~0.8~ & ~~1.0~ & $588 \times 336 \times 175$ & $42.0 \times 24.0 \times 7.0$ & $56 + 56$\\
       SL3H3   & 25.0 & ~~0.8~ & ~~1.0~ & $616 \times 336 \times 175$ & $44.0 \times 24.0 \times 7.0$ & $132 + 132$\\
       AXL1H1   & 6.25 & ~~1.0~ & ~~1.0~ & $672 \times 336 \times 160$ & $48.0 \times 24.0 \times 7.3$ & $36 + 36$\\
       AXL2H1   & 11.1 & ~~1.0~ & ~~1.0~ & $672 \times 336 \times 160$ & $48.0 \times 24.0 \times 7.3$ & $64 + 64$\\
       AXL3H1   & 25.0 & ~~1.0~ & ~~1.0~ & $616 \times 336 \times 160$ & $44.0 \times 24.0 \times 7.3$ & $132 + 132$\\
       AXL1H2   & 6.25 & ~~1.2~ & ~~1.0~ & $672 \times 384 \times 180$ & $56.0 \times 32.0 \times 9.0$ & $56 + 56$\\
       AXL2H2   & 11.1 & ~~1.2~ & ~~1.0~ & $648 \times 360 \times 180$ & $54.0 \times 30.0 \times 9.0$ & $90 + 90$\\
       AXL3H2   & 25.0 & ~~1.2~ & ~~1.0~ & $672 \times 336 \times 180$ & $56.0 \times 28.0 \times 9.0$ & $196 + 196$\\
       AXL1H3   & 6.25 & ~~0.8~ & ~~1.0~ & $672 \times 336 \times 175$ & $48.0 \times 24.0 \times 7.0$ & $36 + 36$\\
       AXL2H3   & 11.1 & ~~0.8~ & ~~1.0~ & $588 \times 336 \times 175$ & $42.0 \times 24.0 \times 7.0$ & $56 + 56$\\
       AXL3H3   & 25.0 & ~~0.8~ & ~~1.0~ & $616 \times 336 \times 175$ & $44.0 \times 24.0 \times 7.0$ & $132 + 132$\\
       AXYL1H1   & 6.25 & ~~1.0~ & ~~1.0~ & $724 \times 362 \times 160$ & $45.3 \times 22.6 \times 7.3$ & $32 + 32$\\
       AXYL2H1   & 11.1 & ~~1.0~ & ~~1.0~ & $916 \times 458 \times 160$ & $50.9 \times 25.5 \times 7.3$ & $72 + 72$\\
       AXYL3H1   & 25.0 & ~~1.0~ & ~~1.0~ & $724 \times 362 \times 160$ & $45.3 \times 22.6 \times 7.3$ & $128 + 128$\\
       SXYL1H1   & 6.25 & ~~1.0~ & ~~1.0~ & $724 \times 362 \times 160$ & $45.3 \times 22.6 \times 7.3$ & $32 + 32$\\
       SXYL2H1   & 11.1 & ~~1.0~ & ~~1.0~ & $916 \times 458 \times 160$ & $50.9 \times 25.5 \times 7.3$ & $72 + 72$\\
       SXYL3H1   & 25.0 & ~~1.0~ & ~~1.0~ & $724 \times 362 \times 160$ & $45.3 \times 22.6 \times 7.3$ & $128 + 128$\\
       NVL1H1   & 3.13 & ~~1.0~ & ~~0.0~ & $576 \times 288 \times 160$ & $48.0 \times 24.0 \times 7.3$ & $36 + 0$\\
       NVL2H1   & 5.55 & ~~1.0~ & ~~0.0~ & $576 \times 288 \times 160$ & $48.0 \times 24.0 \times 7.3$ & $64 + 0$\\
       NVL3H1   & 12.5 & ~~1.0~ & ~~0.0~ & $528 \times 288 \times 160$ & $44.0 \times 24.0 \times 7.3$ & $132 + 0$\\
  \end{tabular}
  \caption{Details of DNS case-setup. $\lambda_p$, $h_1/b$ and $h_2/b$ denote planar packing density, peak-height-to-width and valley-depth-to-width ratio respectively. $N_{x,y,z}$ denotes number of grid cells in streamwise($x$), spanwise($y$) and wall-normal($z$) directions. N denotes the number of peaks plus valleys in the case. 
  Note that the surface coverage densities of the NV cases are half the value of their AX (and AY) counterparts.}
  \label{tab:Cases_setup}
  \end{center}
\end{table}

The packing density $\lambda_p$ in \eqref{eq:12.1} is defined in the same way as plan solidity in \cite{Chung2021PredictingSurfaces}. $A_p$ comprises the plan area of roughness elements, which in this case includes the base area of peaks and valleys.

Additional constraints have been placed to ensure that the grid surface matches the cube surface for accurate roughness representation in the simulations. 
This is done by enforcing $b/\Delta_x$, $b/\Delta_y$, $h_1/\Delta_z$ and $h_2/\Delta_z$ to be integers where $\Delta_x$, $\Delta_y$ and $\Delta_z$ stand for the grid spacing in the streamwise, spanwise and wall-normal directions. This has been implemented for all cases except the $45^\circ$ alignment rough surfaces. 

\begin{figure}
  \centerline{\includegraphics[width=1.0\linewidth]{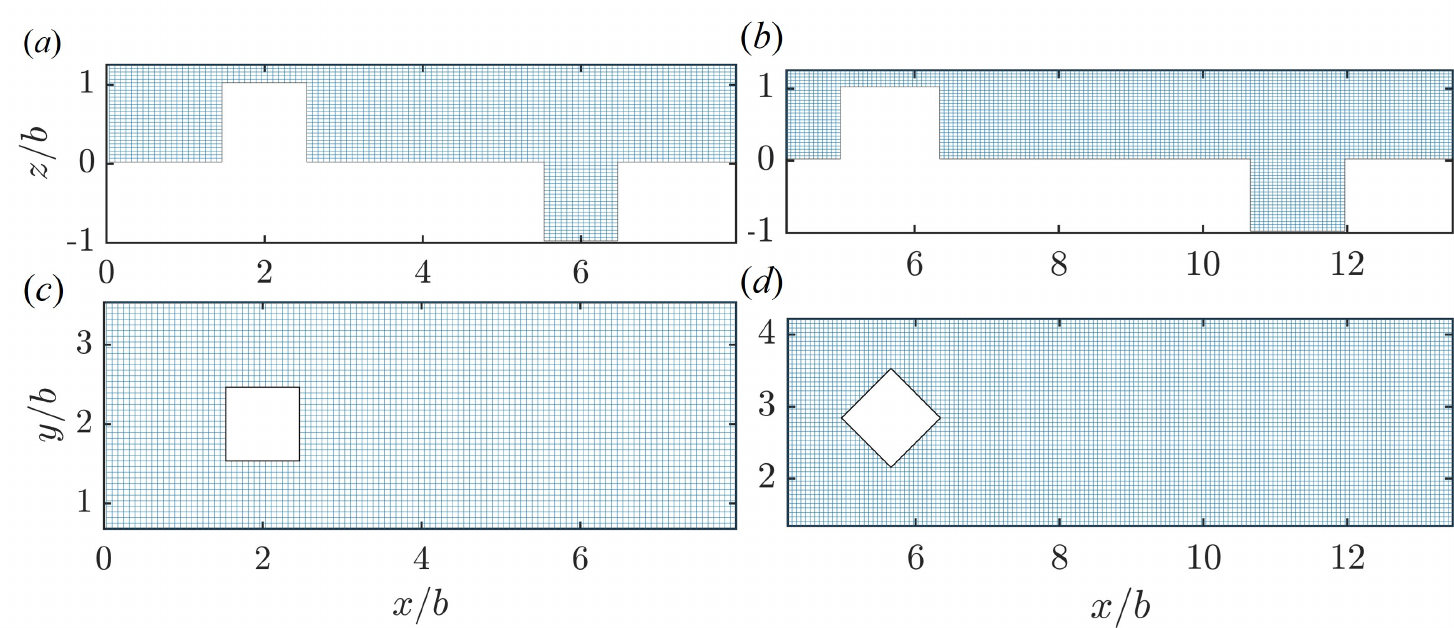}}
  \caption{Top and side views of DNS grids used for two cases: AXL1H1(\textit{a,c}) and AXYL1H1(\textit{b,d}). The side views are streamwise wall-normal ($xz$) planes cut at the middle of the cubes. The top views are wall-parallel ($xy$) planes cut near the roughness peak.}
\label{fig:grid}
\end{figure}

As can be seen from figure \ref{fig:grid}, the grid is in perfect alignment with the roughness element in AYL1H1 as opposed to AXYL1H1. 
As a result, there are minor variations in the average roughness height $k_a/b$, root-mean-square roughness height $k_{rms}/k_a$ and kurtosis $Ku$ for AXYLiH1 and SXYLiH1 surfaces when compared with AYLiH1 and SLiH1 where i=1, 2, 3. 
This can be noticed in Table \ref{tab:DNS results}. 
A finer resolution has been opted for in all such cases to minimize these variations
The streamwise and spanwise grid resolutions, in terms of number of cells per roughness element, are given by $b/\Delta_x$ = $b/\Delta_y $ = 12-14 for AX, AY, S and NV surfaces. These grid resolutions are about 16-18 for AXY and SXY surfaces.
The wall-normal grid resolution is in the range of $b/\Delta_z$ = 20-25 for all cases.

\section{DNS results}\label{sec:3}

The rough walls we study cover an underexplored region in roughness parameter space and therefore are a good addition to the rough wall literature.
In this section, we first discuss qualitative findings pertaining to the instantaneous flow-field.
Subsequently, the quality of our channel flow simulations is examined with the help of the mean momentum budget, followed by results of mean velocity profiles and quantitative estimates of effective roughness height ($z_0$) and zero-plane displacement height ($d$).
The section concludes with Reynolds and dispersive stress results and comparisons for all 36 cases.

\subsection{Instantaneous flow field}

\begin{figure}
 \centerline{\includegraphics[width=1.0\linewidth]{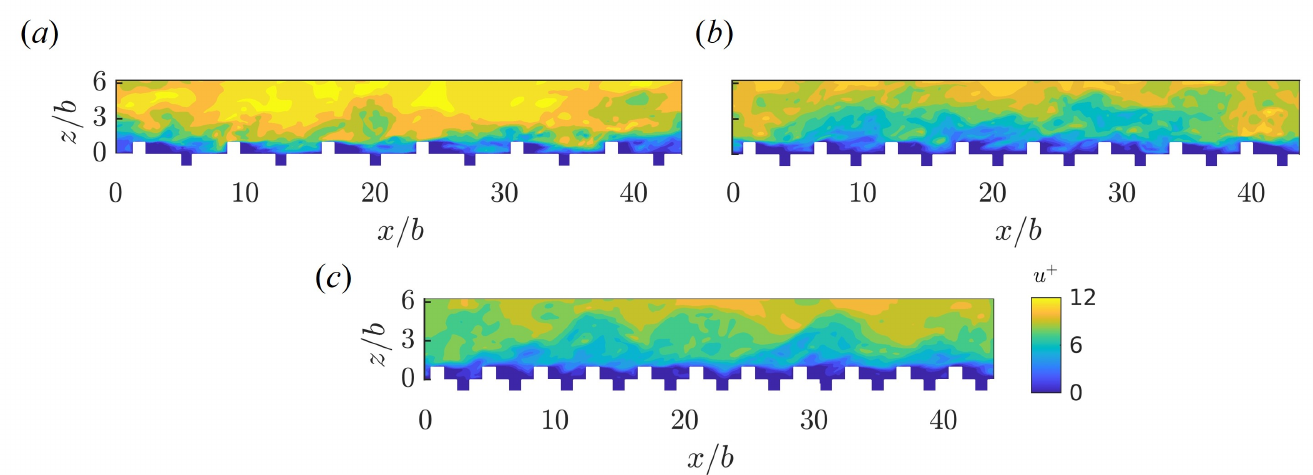}}
  \caption{Streamwise instantaneous velocity contours at $y/L_y \approx 0.43$ for (\textit{a}) AYL1H1, (\textit{b}) AYL2H1 and (\textit{c}) AYL3H1.}
\label{fig:Inst_vel_wall_normal}
\end{figure}

We begin by presenting the instantaneous flow field as an introduction to the dataset.
From the contours in figure \ref{fig:Inst_vel_wall_normal}, an overall decrease in instantaneous bulk-velocity can be observed with increasing $\lambda_p$. 
A region of reduced streamwise velocity can be noted in the immediate wake of the protrusions.
The flow in the pits can be observed to be less energetic and their interactions with the outer layer are minimal.

\begin{figure}  \centerline{\includegraphics[width=0.9\linewidth]{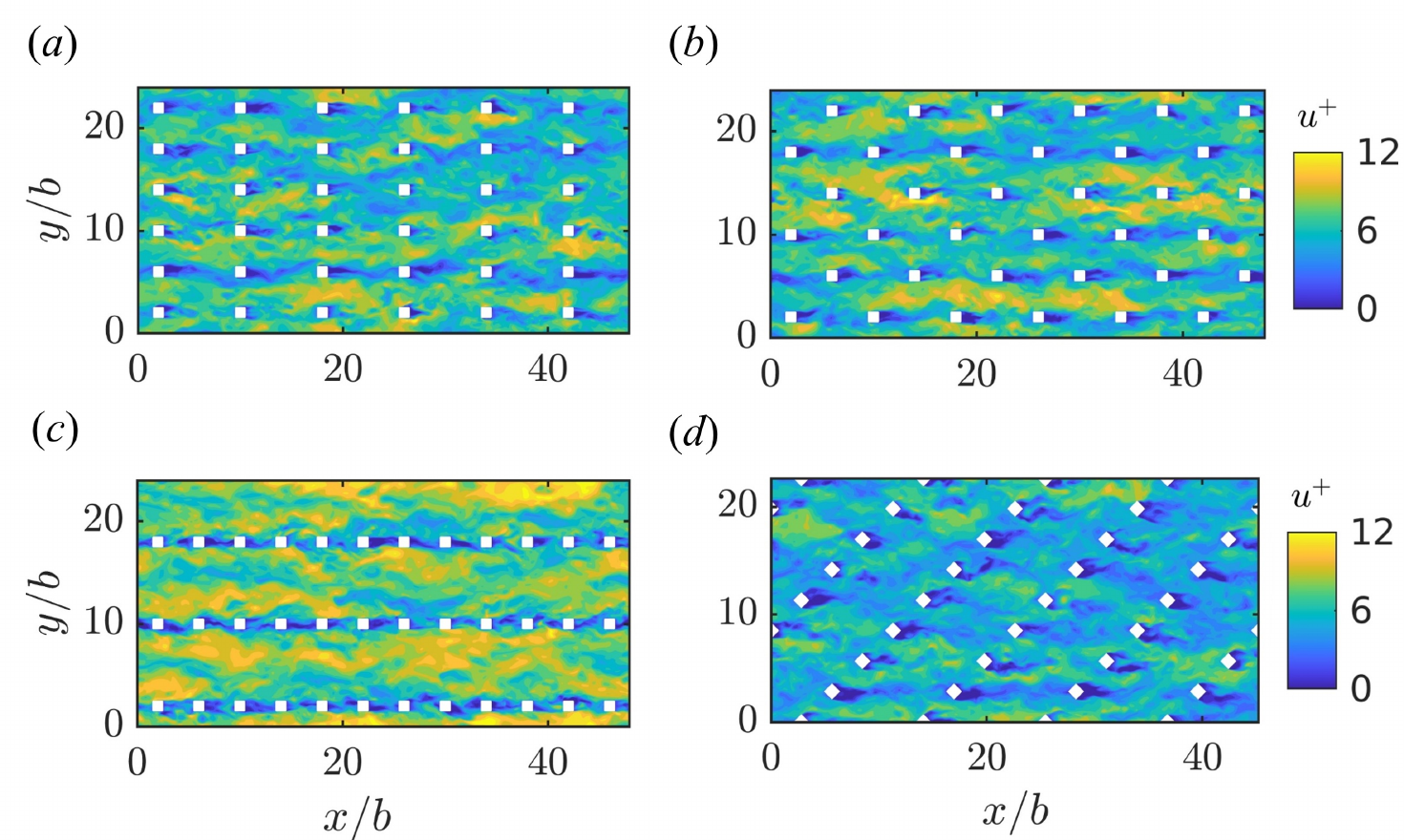}}
  \caption{Streamwise instantaneous velocity contours at $z/b \approx 1.0$ for (\textit{a}) AYL1H1, (\textit{b}) SL1H1, (\textit{c}) AXL1H1  and (\textit{d}) AXYL1H1.}
\label{fig:Inst_vel_wall_parallel}
\end{figure}

Figure \ref{fig:Inst_vel_wall_parallel} shows the velocity contours on a wall-parallel plane near the peak height for different roughness element arrangements at the same $\lambda_p$. 
By visual inspection, contours in \ref{fig:Inst_vel_wall_parallel}(\textit{a}) and \ref{fig:Inst_vel_wall_parallel}(\textit{b}) are observed to have similar and intermediate average streamwise velocities. \ref{fig:Inst_vel_wall_parallel}(\textit{c}) contains the highest and \ref{fig:Inst_vel_wall_parallel}(\textit{d}) the lowest average streamwise velocities amongst these four cases.
In all cases, streamwise streaks of high and low velocities can be observed near the roughness peak.
These streaks are noticeably longer in certain cases, such as in figure \ref{fig:Inst_vel_wall_parallel}(\textit{c}), as the wakes overlap each other.
These wake interactions become more pronounced at higher $\lambda_p$ and result in a flow-sheltering effect as observed in densely packed surfaces such as in \cite{xu2021flow}.

Although we will pursue a quantitative analysis in the next subsections, here we may already conclude first that the valleys do not contribute significantly to the drag, and second that two-point rough wall statistics are important to the characterization of the equivalent sand-grain roughness height $k_s$.

\subsection{Statistical convergence}\label{sec:3.2}

Before presenting the flow statistics, we first check the statistical convergence of our data.
The statistical convergence can be evaluated with the help of the mean momentum budget following \cite{coceal2006mean}, \cite{leonardi2010channel}, \cite{xu2021flow} and  \cite{Zhang2022EvidenceFlows}.
The equation is comprised of viscous diffusion, turbulent transport, dispersive stress and pressure-gradient terms as shown below:
\begin{equation}\label{eq:budget}
\nu \frac{\text{d}\langle \overline{u} \rangle}{\text{d}z} - \langle \overline{u^\prime w^\prime} \rangle - \langle u^{\prime \prime}w^{\prime \prime} \rangle = \frac{1}{\rho} \frac{\text{d} \langle\overline{p}\rangle}{\text{d}x}(z-\delta)
\end{equation}
It is considered to be statistically converged when the sum of these stresses is a linear function of the wall-normal distance ($z$), i.e., when the budget balances without the unsteady term.
Note that \eqref{eq:budget} is obtained by integrating the x-component of the double-averaged momentum equation:
\begin{equation}\label{eq:budget2}
\frac{\text{d}}{\text{d}z}\left(\nu \frac{\text{d}\langle \overline{u} \rangle}{\text{d}z} - \langle \overline{u^\prime w^\prime} \rangle - \langle u^{\prime \prime}w^{\prime \prime} \rangle \right) - \frac{1}{\rho} \frac{\text{d} \langle\overline{p}\rangle}{\text{d}x} - f = 0
\end{equation}
found in equation 15 of \cite{nikora2013spatially}, with static roughness in place of the mobile roughness assumed by the authors. Here $f$ represents the drag force. 
As $f$ is non-zero within the roughness occupied region, \eqref{eq:budget} applies to the region outside the roughness only.
The averaging procedure corresponds to the double-averaging method, common in roughness literature, introduced by \cite{raupach1982averaging} for flows within vegetation canopies.  
The spatial averaging used in the context of \eqref{eq:budget} and \eqref{eq:budget2} is superficial, as per the definition for superficial averaging in \cite{schmid2019volume}.

\begin{figure}
  \centerline{\includegraphics[width=1\linewidth]{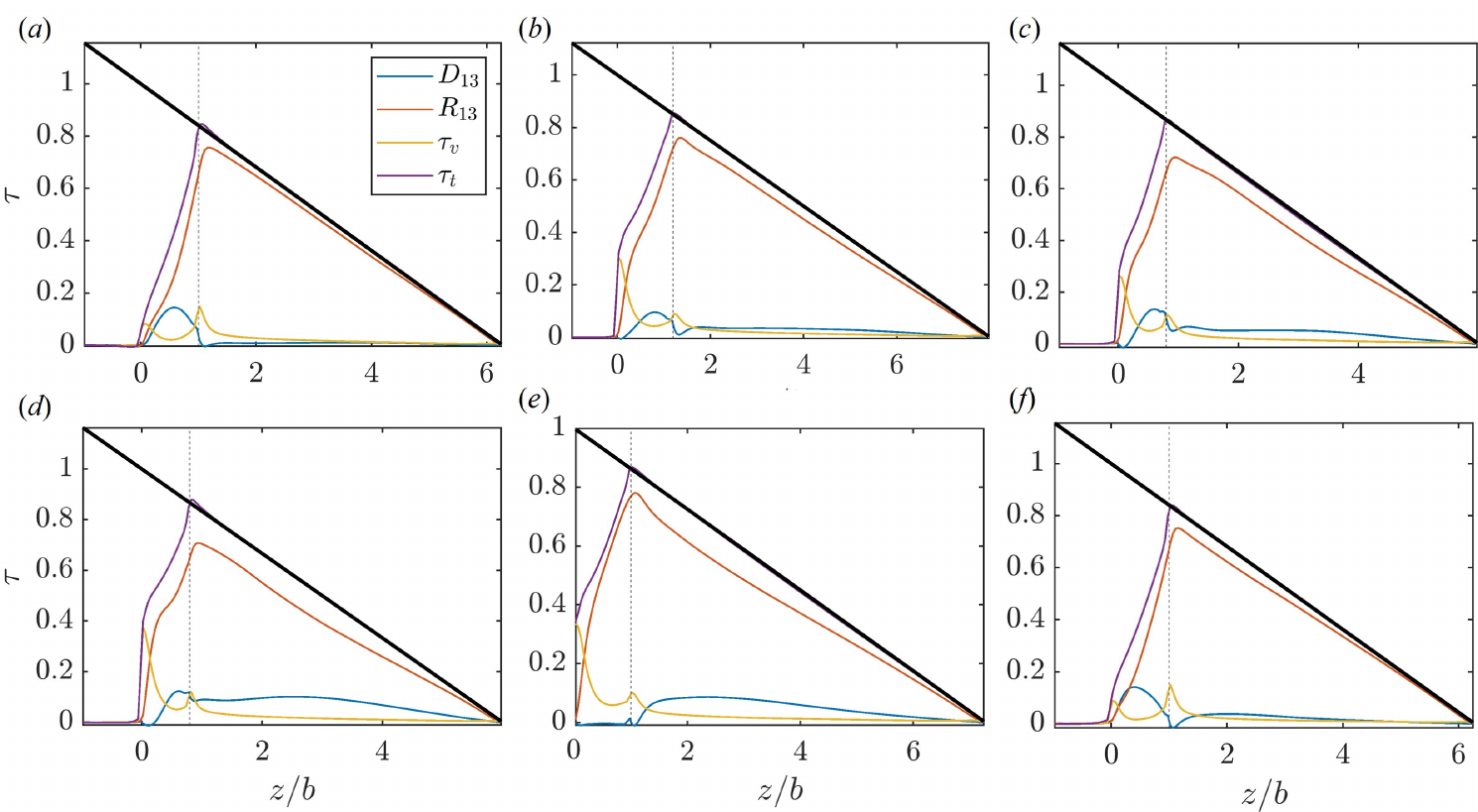}}
  \caption{The stress budget plot comprising Reynolds stress $R_{13}$, dispersive stress $D_{13}$, viscous stress $\tau_v$ and total stress $\tau_t$ for cases: (\textit{a}) AYL3H1, (\textit{b}) SL1H2, (\textit{c}) SL2H3, (\textit{d}) AXL2H3, (\textit{e}) NVL1H1 and (\textit{f}) AXYL3H1. The black solid line corresponds to $1-z/\delta$.The vertical dashed line denotes the peak roughness height.}
\label{fig:budget plot}
\end{figure}

Figure \ref{fig:budget plot} shows these terms for a few cases. 
These cases have been selected so that they include different $\lambda_p$, $Sk$ and roughness arrangements while maintaining brevity.
We see that the total stresses in all cases follow a linear function of $z/b$, and therefore our DNSs are statistically converged.
In addition, we make the following observations.
The Reynolds stress $R_{13}$ can be observed to be the largest in magnitude as compared to the viscous stress $\tau_v$ and dispersive stress $D_{13}$ outside the roughness occupied layer. 
The roughness occupied layer is marked by the vertical line in figure \ref{fig:budget plot}.
In the roughness-occupied layer, i.e. below this marking, $\tau_v$ is observed to contain two maxima, one caused by the surface at $z=0$ and the other near the cube height $z=h_1$.
The latter is also observed in other DNS studies by \cite{xu2021flow} and \cite{Zhang2022EvidenceFlows}.
Meanwhile, $R_{13}$ attains a maximum just above the peak height and declines to zero as $z$ approach the outer layer.
$D_{13}$ is small and contained within the roughness region for most cases.
For $z/b < 0$, i.e. below the surface in the region of valleys, all stresses are observed to be negligible.
Note that the figure \ref{fig:budget plot}(e) starts from $z/b = 0$ as the case NVL1H1 contains no valleys.

\begin{figure}
  \centerline{\includegraphics[width=1.0\linewidth]{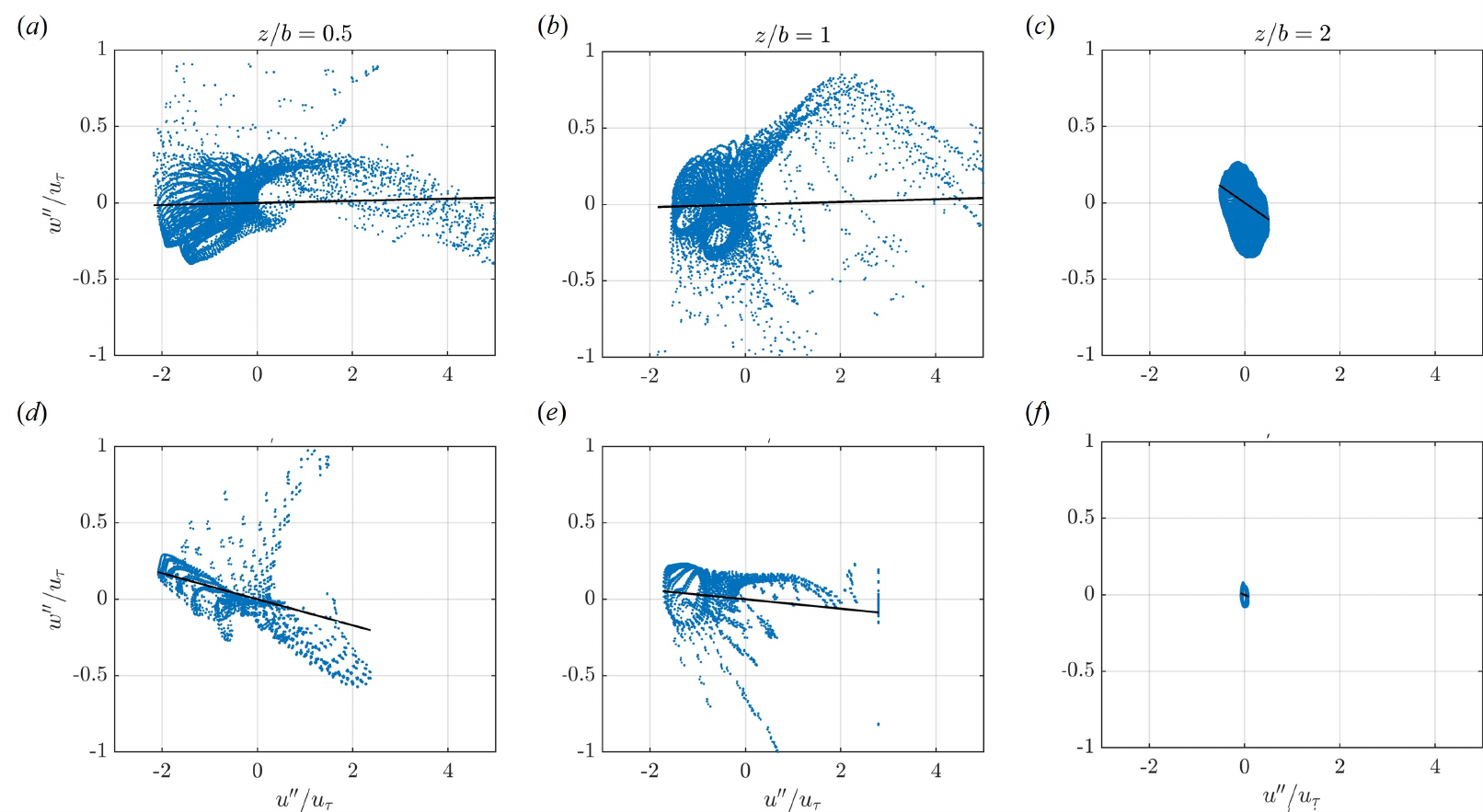}}
  \caption{Quadrant analysis for the dispersive stress components of $D_{13}$ for  cases NVL1H1 at   
  (\textit{a}) $z/b=0.5$, (\textit{b}) $z/b=1$, (\textit{c}) $z/b=2$ and AYL3H1 at (\textit{d}) $z/b=0.5$, (\textit{e}) $z/b=1$ and (\textit{f}) $z/b=2$. The solid black line depicts a linear fit on the data points.
}
\label{fig:D13_exp}
\end{figure}

It might be worth noting certain differences in dispersive stress $D_{13}$ in figure \ref{fig:budget plot}. 
For instance, most noticeable would be the differences in cases AYL3H1 and NVL1H1.
$D_{13}$ is observed to be non-negligible within the roughness occupied region for AYL3H1 whereas $D_{13}$ for NVL1H1 inside the sublayer is almost zero.
Figure \ref{fig:D13_exp} tries to provide an intuitive explanation to this.
Here, the dispersive streamwise and wall-normal fluctuating components, $u^{\prime \prime}$ and $w^{\prime \prime}$, are shown for averaged repeating tiles for cases NVL1H1 and AYL3H1.
Three planes, one within, the others at and above the roughness crest are chosen to display their scatter. 
When $u^{\prime \prime}$ and $w^{\prime \prime}$ predominantly lie on the second or fourth quadrant, the dispersive stress $-\langle u^{\prime \prime}w^{\prime \prime}\rangle$ becomes positive.  
The negative slope of the linear fit is also indicative of this.
This can be seen in figures \ref{fig:D13_exp}(c), which correspond to NVL1H1 just above the roughness and \ref{fig:D13_exp} (d), which correspond to AYL3H1 within the roughness.
These may indirectly imply mean flow inhomogeneity, which is due to the presence of surface roughness and/or secondary flows.
On the other hand, when the points are more evenly spread across all quadrants, as is the case in figures \ref{fig:D13_exp}(a), \ref{fig:D13_exp}(b) and \ref{fig:D13_exp}(e) or too close to the origin as in \ref{fig:D13_exp}(f), this dispersive stress becomes negliglible. 

\subsection{Mean velocity profiles}

The mean flow statistics are presented in this subsection.
Additionally, the relevant rough surface parameters and the hydrodynamic properties determined from mean velocity profiles are also listed in Table \ref{tab:DNS results}.

\begin{figure}  
\centerline{\includegraphics[width=0.8\linewidth]{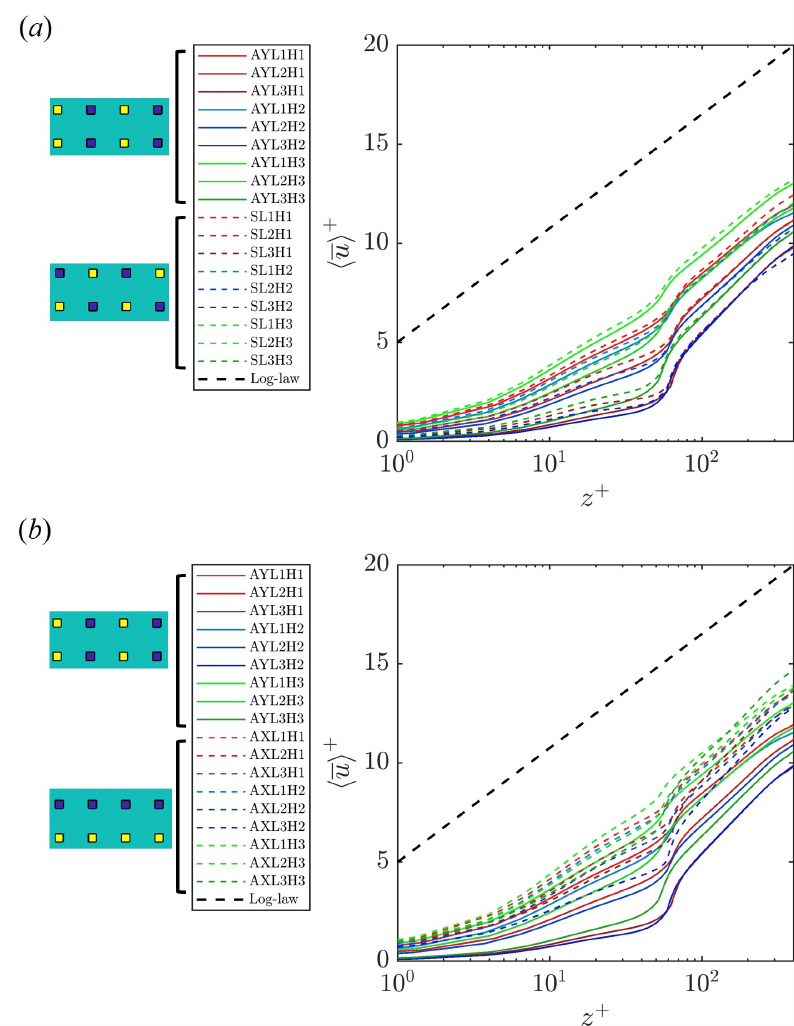}}
\caption{Mean streamwise velocity profile comparing (\textit{a}) Spanwise-aligned and staggered cases and (\textit{b}) spanwise-aligned and streamwise-aligned cases.}
\label{fig:Al Vs St}
\end{figure}

Figure \ref{fig:Al Vs St}(a) shows the inner scaled mean velocity profiles, comparing the spanwise aligned and staggered cases, i.e., the AY cases and the S cases. 
The following observations can be duly noted. 
First, the higher $\lambda_p$ cases of surface coverage density 25\%, comprising AYL3H1, AYL3H2, AYL3H3 and their staggered counterparts SL3H1, SL3H2 and SL3H3, produce lower magnitudes of mean-velocity.
This is expected for roughness in the k-type regime.
Second, the staggered arrangement produces similar velocity profiles to the spanwise aligned arrangement. 
This is not unexpected, and similar observations were made in \cite{yang2016exponential}.
Due to a large spanwise distance between neighbouring roughness elements in the spanwise direction, staggering the roughness element in the streamwise direction has little effect on the mean flow.
Lastly, at constant $\lambda_p$, cases with higher skewness ($Sk$) produce lower mean-velocity profiles.
This role of $Sk$ is apparent as we notice increased drag with increased $Sk$ (consistent with systematic studies performed by \cite{flack2020skin}) and we observe this effect in all our cases irrespective of arrangement. 
For example, comparing (AYL1H1, AYL1H2 and AYL1H3) from figure \ref{fig:Al Vs St}(a) or (AXL1H1, AXL1H2 and AXL1H3) from figure \ref{fig:Al Vs St}(b) at the same $z^+$, one can notice that the mean velocities in the log layer are slightly lower for the surface with higher $Sk$ (H3 cases being lower than H2 and H1 cases).
In the present dataset, variations in $Sk$ are introduced by varying the height of the peaks, while maintaining the depth of the dips constant, and therefore any changes due to $Sk$ could also be viewed as due to the roughness height.
Similarly, as variations in $k_{\rm rms}/k_a$ are introduced by varying the surface coverage density, any changes due to $k_{\rm rms}/k_a$ could also be viewed as due to surface coverage density.

Similarly, figure \ref{fig:Al Vs St}(b) compares mean velocity profiles for roughness with streamwise and spanwise alignment, i.e., AX cases and AY cases.
When the roughness features are aligned in the streamwise direction, higher mean velocities can be observed as compared to those aligned in the spanwise direction.
An intuitive explanation for this is the unobstructed passage of fluid between two rows of roughness elements giving the fluid less drag when they are aligned in the streamwise direction. 
This is also evident from the instantaneous velocity contours in figure \ref{fig:Inst_vel_wall_parallel}(c).

\begin{figure}
\centerline{\includegraphics[width=0.95\linewidth]{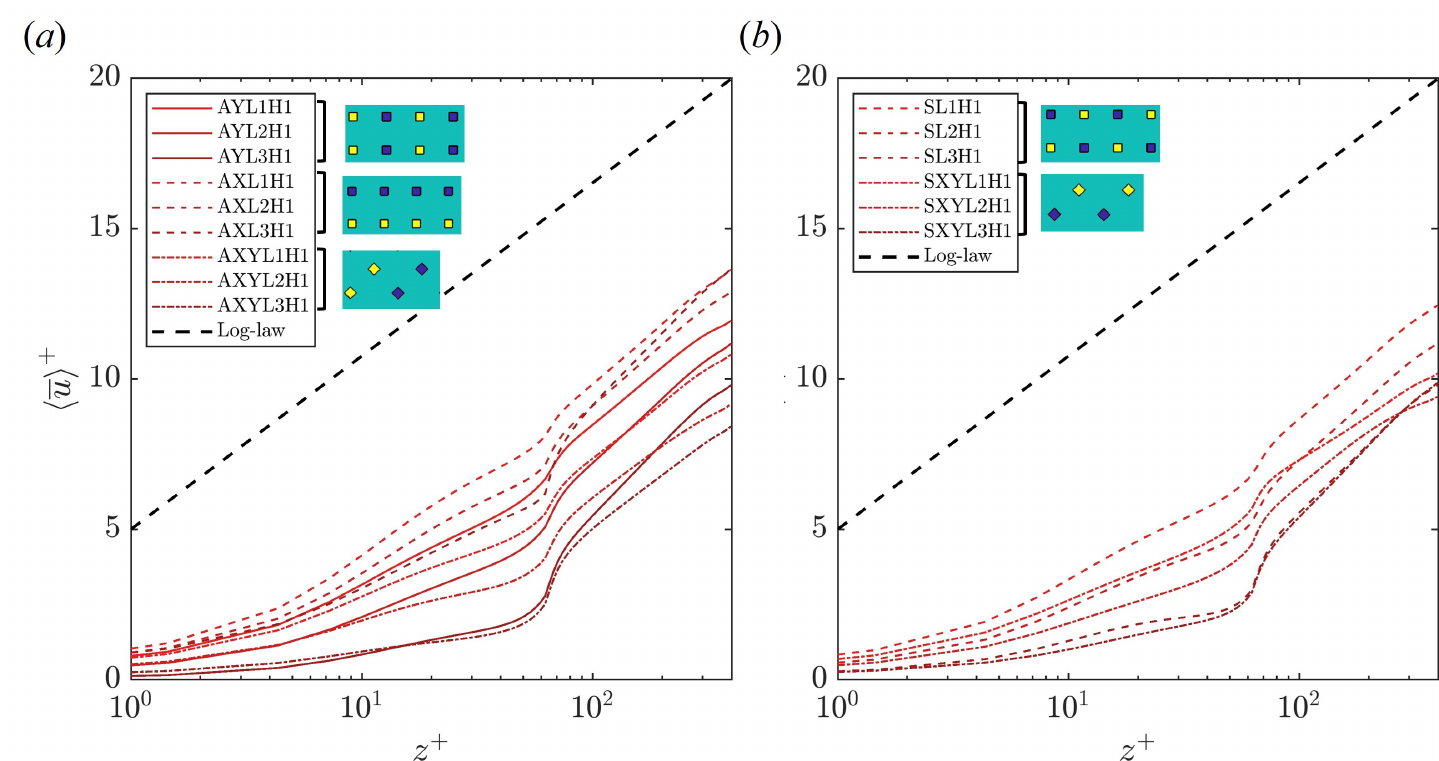}}
  \caption{Mean streamwise velocity profile comparing (\textit{a}) spanwise, streamwise and $45^\circ$ aligned cases and (\textit{b}) staggered and staggered $45^\circ$ cases.}
\label{fig:VelocityprofileASXY}
\end{figure}

Figure \ref{fig:VelocityprofileASXY} includes the mean velocity profiles from the $45^\circ$ arrangements (XY), with figure \ref{fig:VelocityprofileASXY}(a) comparing AY, AX, AXY and figure \ref{fig:VelocityprofileASXY}(b) comparing S and SXY. 
For the same $\lambda_p$, the surfaces follow AX $>$ AY $>$AXY in (a) and S $>$ SXY in (b) for the order of mean velocity magnitudes in the log-law region. 
The orientation of the roughness element is found to be an important factor here. 
AXY and SXY surfaces possess a higher frontal area (i.e. higher solidity) as compared to their counterparts AX, AY, or S, which is the reason for the higher drag observed for these surfaces.

\begin{figure}
  \centerline{\includegraphics[width=0.9\linewidth]{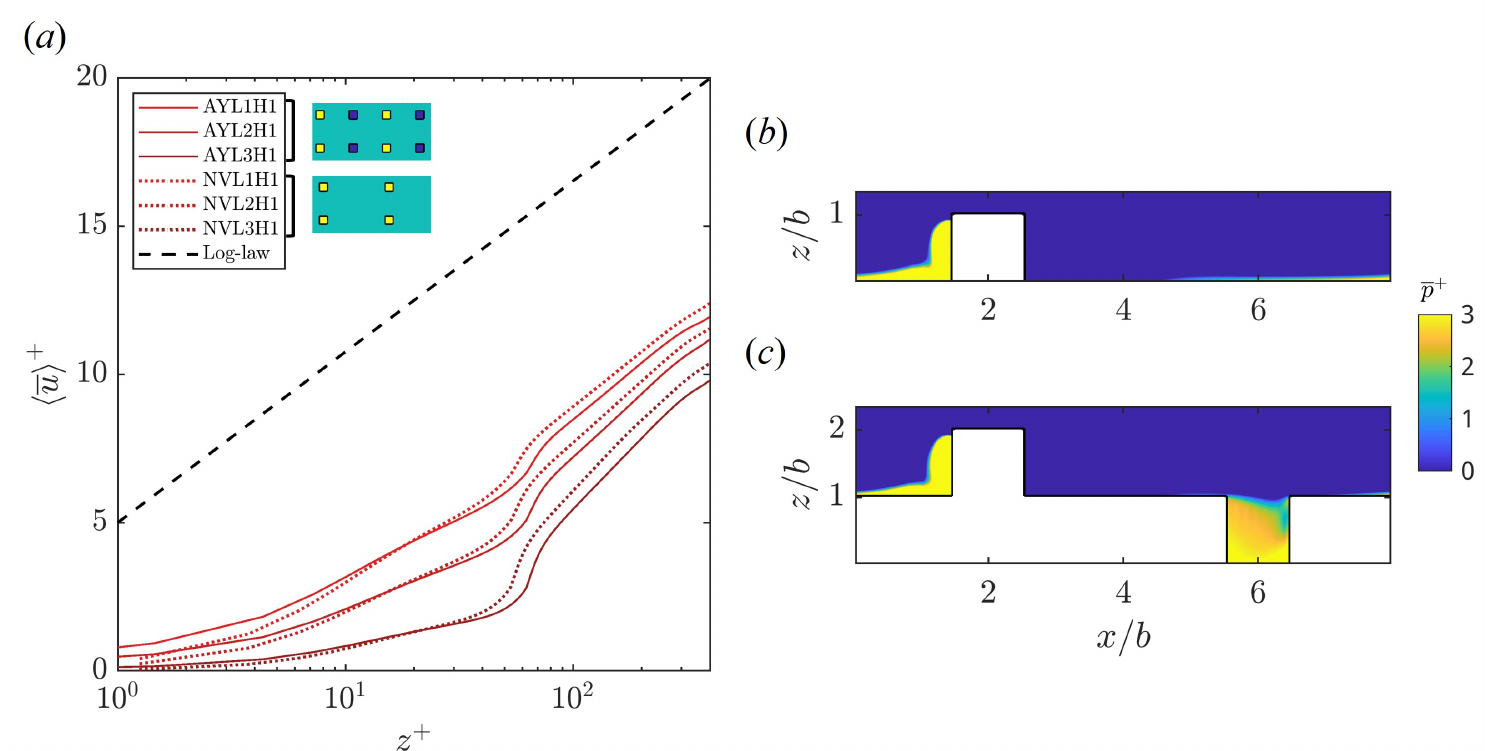}}
  \caption{Details depicting the effect of valleys: (\textit{a}) Mean streamwise velocity profile comparing cases with and without valleys.  The solid and dashed lines depict spanwise-aligned and no-valleys cases respectively. (\textit{b}) Mean pressure contours for a repeating tile in NVL1H1 and (\textit{c}) Mean pressure contours for a repeating tile in AYL1H1.}
\label{fig:Eff_valley}
\end{figure}

Figure \ref{fig:Eff_valley}(a) shows the mean velocity profiles comparing three pairs of surfaces with and without valleys. 
It can be observed that the valleys do contribute to mean velocity reduction but only by a small amount.
This reduced mean velocity can be inferred to be caused by the additional pressure drag contributions from the valleys.
In support of this claim, 
figures \ref{fig:Eff_valley}(b) and \ref{fig:Eff_valley}(c) are shown, which contain mean pressure contours that are averaged over repeating tiles in both $(x, y)-$ directions in AYL1H1 and NVL1H1.
In these pressure contours, which are normalized by $\rho u_{\tau}^2$, the volume-averaged mean pressure is taken as the reference pressure. 
The additional pressure drop can be clearly noticed in valleys.
Note that the observation here cannot be generalized to all dips. 
In particular, dips or valleys will play a significant role if they occupy a considerable part of the surface area, in which scenario, a new bottom surface forms and the protrusions between the dips can be viewed as surface roughness.

Next, we measure the roughness' hydrodynamic properties.
The effective roughness height $z_0$ and zero-plane displacement height $d$ for the rough surfaces are determined from the mean streamwise velocity profiles as follows
\begin{equation}\label{eq:4}
\frac{\langle \overline{u} \rangle}{u_\tau}=\frac{1}{\kappa} \log \frac{z-d}{z_0} 
\end{equation}
where $\langle \overline{u} \rangle$ and $z$ in \eqref{eq:4} are taken from the log-law region, which has been taken to be in the range $(z-h_1)^+ \approx 30-100$. Further evidence to the existence of log-law in this range will be discussed later with the help of figure \ref{fig:log-law collapse}. 
Linear regression of the expression \eqref{eq:4} yields the resulting $z_0$ and $d$ reported in Table \ref{tab:DNS results}. 
Here, the von-K\'arm\'an constant is set to $\kappa = 0.4$. 
There can be alternate approaches to calculate $d$  from the centroid of the total force distribution following \cite{jackson1981displacement} and \cite{coceal2006mean}.
However, it is also observed that the centre of the force could sometimes underestimate $d$ when the flow primarily interacts with the top portion of the roughness as seen in figure 7 of \cite{xu2021flow}.
Here, we prefer regression fitting over this approach due to the aforementioned reason.

\begin{figure}
  \centerline{\includegraphics[width=1.0\linewidth]{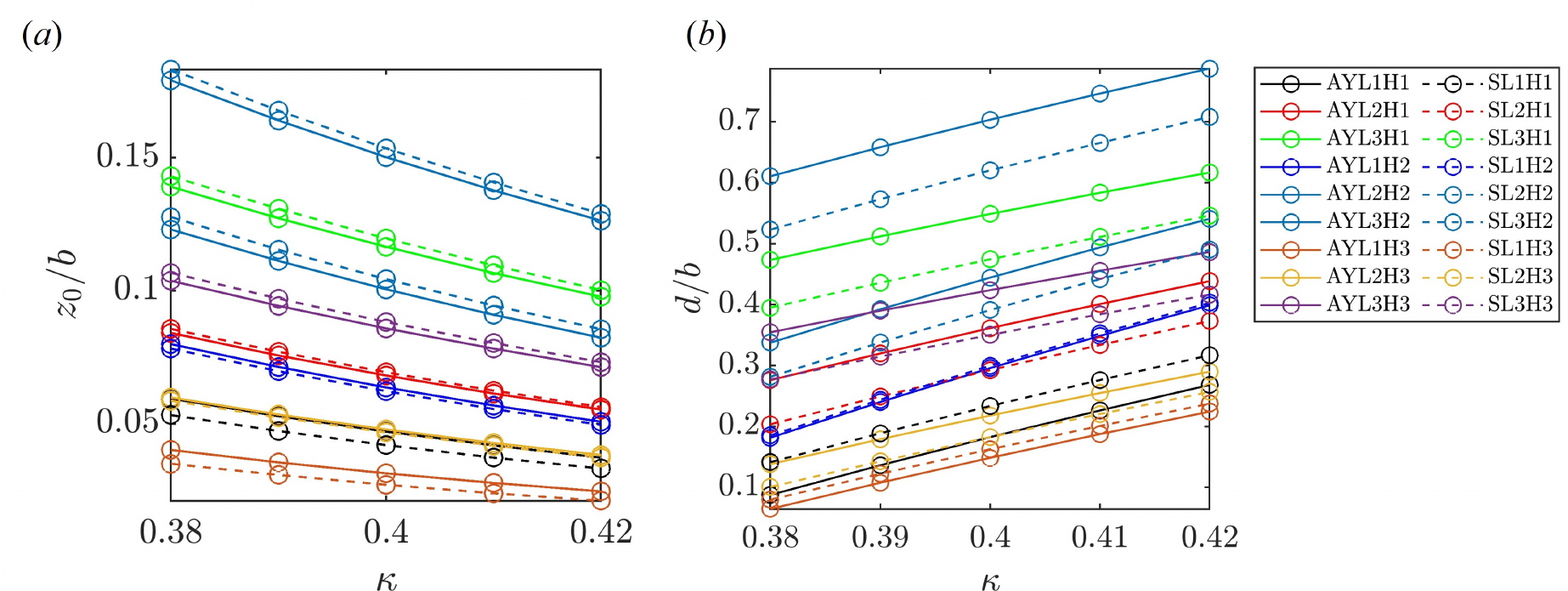}}
  \caption{Plot showing the sensitivity to variations in von-K\'arm\'an constant $\kappa$ for  spanwise aligned (AY) and staggered (S) cases for the calculated (\textit{a}) effective roughness height $z_0/b$ and (\textit{b}) zero-plane displacement height $d/b$.}
\label{fig:z0_uncertainty}
\end{figure}

\begin{figure}
  \centerline{\includegraphics[width=0.95\linewidth]{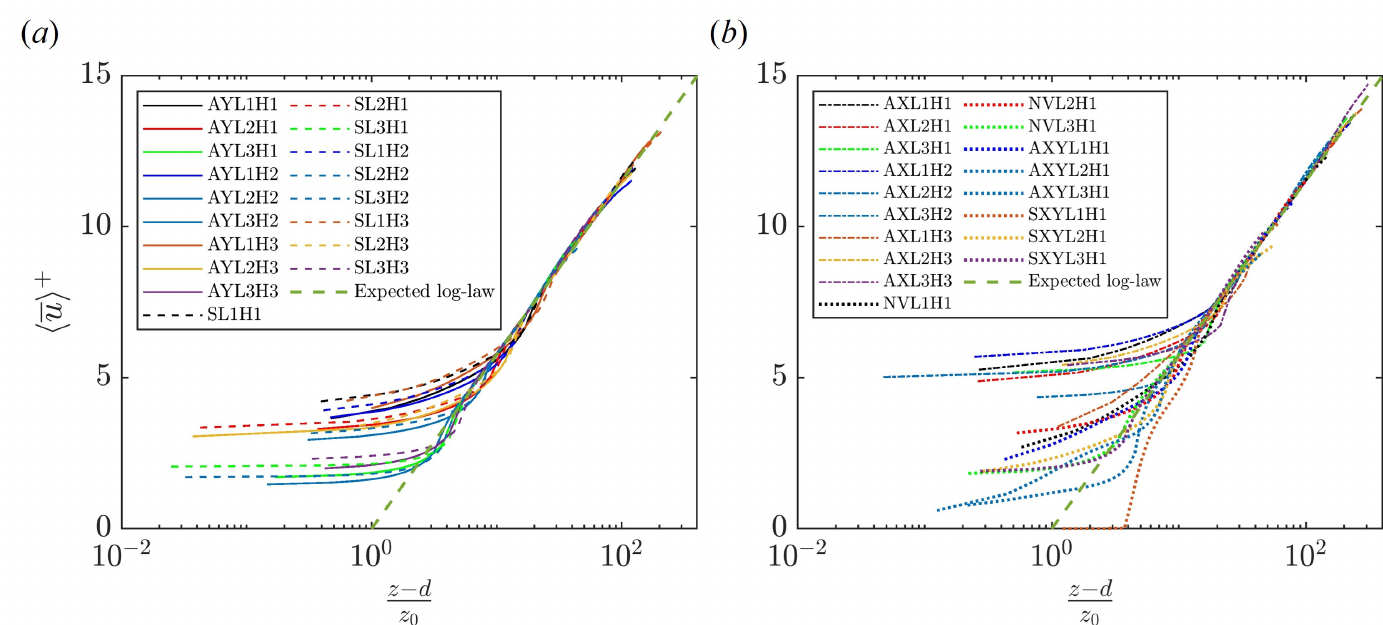}}
  \caption{Plot showing the log-law collapse for all cases: (a) AY, S cases, (b) AX, NV, AXY and SXY cases.}
\label{fig:log-law collapse}
\end{figure}

Since we assume the value of  von-K\'arm\'an constant $\kappa = 0.4$ for our regression procedure, checking the changes in measured $z_0/b$ due to variations in $\kappa$ is important.
Figure \ref{fig:z0_uncertainty} shows the sensitivity of calculated $z_0/b$ and $d/b$ with various values of $\kappa$. 
It can be seen that $z_0/b$ decreases and $d/b$ increases with $\kappa$. 
While we do see noticeable variations in these values, it is worth noting that these changes would not be significant on the log scale, which is the relevant scale for the drag produced by these surfaces.
Also, for a given value of $\kappa$ the differences in $z_0/b$ and $d/b$ are almost the same, which means it is reasonable to compare these values for the different surfaces.

Further, we note that determining the roughness height $z_0$ or equivalent sandgrain roughness height $k_s$ requires the flow to be in the fully rough regime.
Based on \cite{jimenez2004turbulent}, flow is considered fully rough when $k_s^+ > 80$. 
In \cite{AghaeiJouybari2021Data-drivenFlows}, the authors have chosen $k_s^+ > 50$ to be their definition for fully rough regime.
We have verified that all surfaces except AXL1H3 and AXL3H3 satisfy this latter condition, the $k_s^+$ values for which stand at 42.6 and 36 respectively.
However, it is important to note that the exact value of $k_s^+$ for which this transition to fully rough flow happens is unknown and also depends on the type of roughness being considered \cite{flack2010review}.
Even so, we could still measure $\Delta U^+$ and $z_0$ and discuss their variations with respect to the roughness as we are making these comparisions at a fixed Reynolds number ($Re_{\tau}=400$).

\begin{figure}
  \centerline{\includegraphics[width=0.95\linewidth]{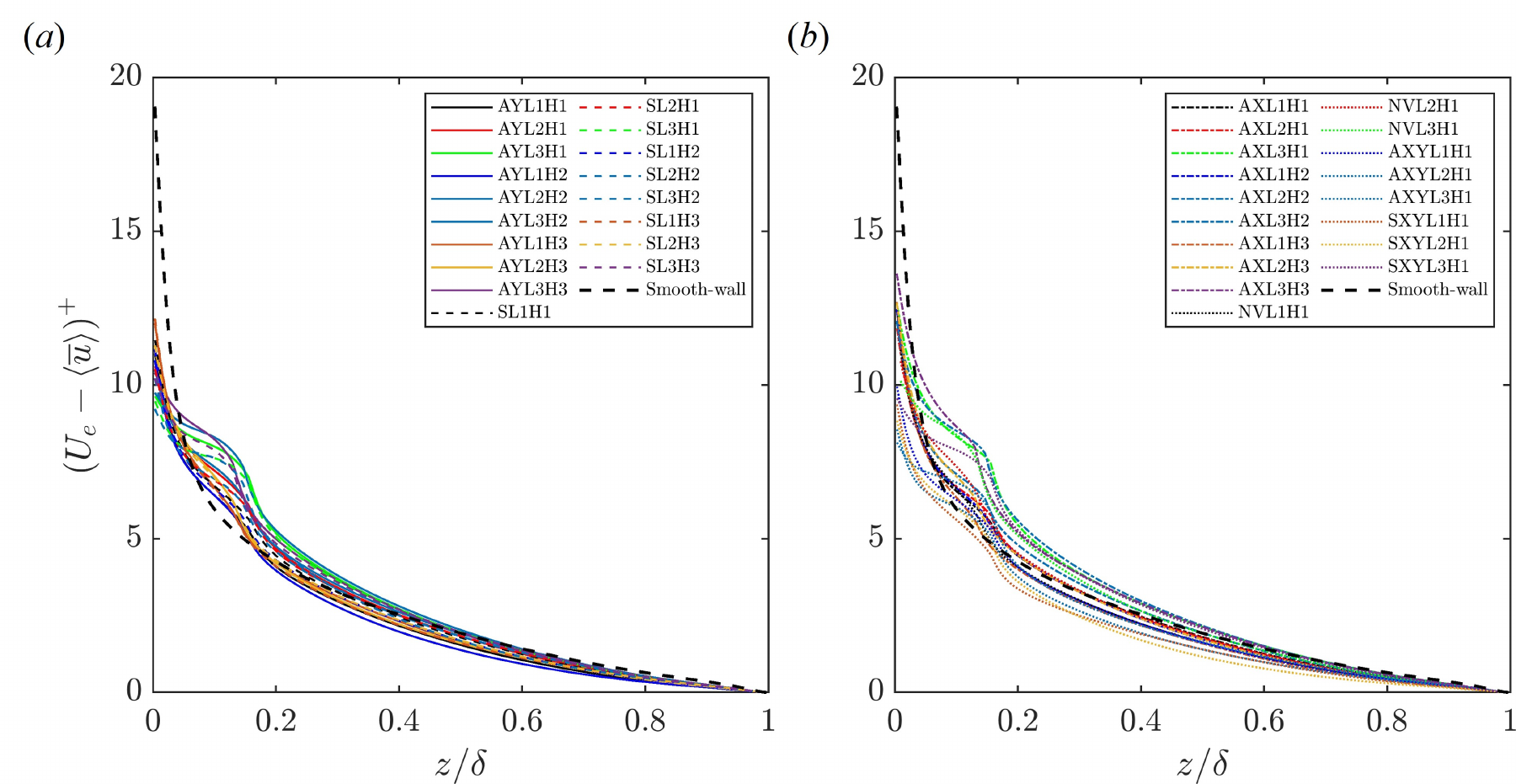}}
  \caption{Velocity defect profiles for all surfaces: (a) AY, S and (b) AX, NV, AXY and SXY cases.}
\label{fig:outer-layer similarity}
\end{figure}

Figure \ref{fig:log-law collapse} illustrates the quality of this log-law fit. 
The mean velocity plots can be observed to show a good collapse in the log layer region for all cases.
The collapse is also a sign of mean-flow universality, wherein such two-parameter forms of mean velocity profiles have been know to be adequate irrespective of the rough surface for $h/\delta \approx 0.03-0.5$. 
Figure \ref{fig:outer-layer similarity} shows mean velocity profiles in defect form.
This is being presented here as evidence for outer-layer similarity.
Reasonably well-collapsed regions can be observed farther away from the wall for the smooth and the rough wall cases, underscoring a universality in mean flow behaviour in the outer layer irrespective of surface conditions for fully rough flow.

\begin{figure}
  \centerline{\includegraphics[width=0.95\linewidth]{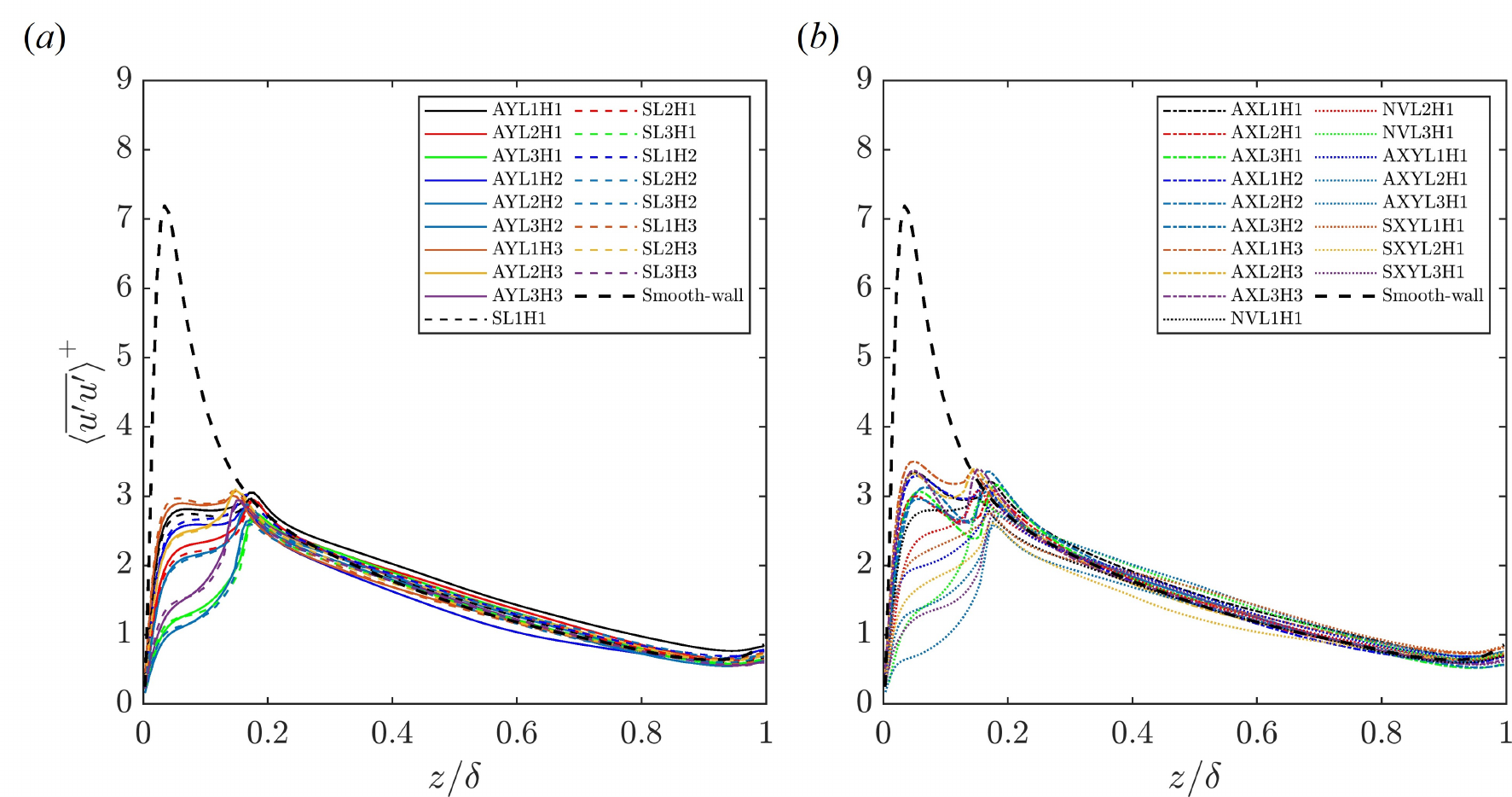}}
  \caption{Normalized streamwise Reynolds stress profiles for all surfaces: (a) AY, S and (b) AX, NV, AXY and SXY cases.}
\label{fig:outer-layer similarity_UU}
\end{figure}

The mean velocity profile for the smooth wall in figures \ref{fig:outer-layer similarity}(a) and (b) was obtained from a DNS of a half-channel with $Re_{\tau} = 400$.
Similarly, figures \ref{fig:outer-layer similarity_UU} depicts outer-layer similarity in the streamwise Reynolds stress profiles.

\begin{figure} 
  \centerline{\includegraphics[width=0.9\linewidth]{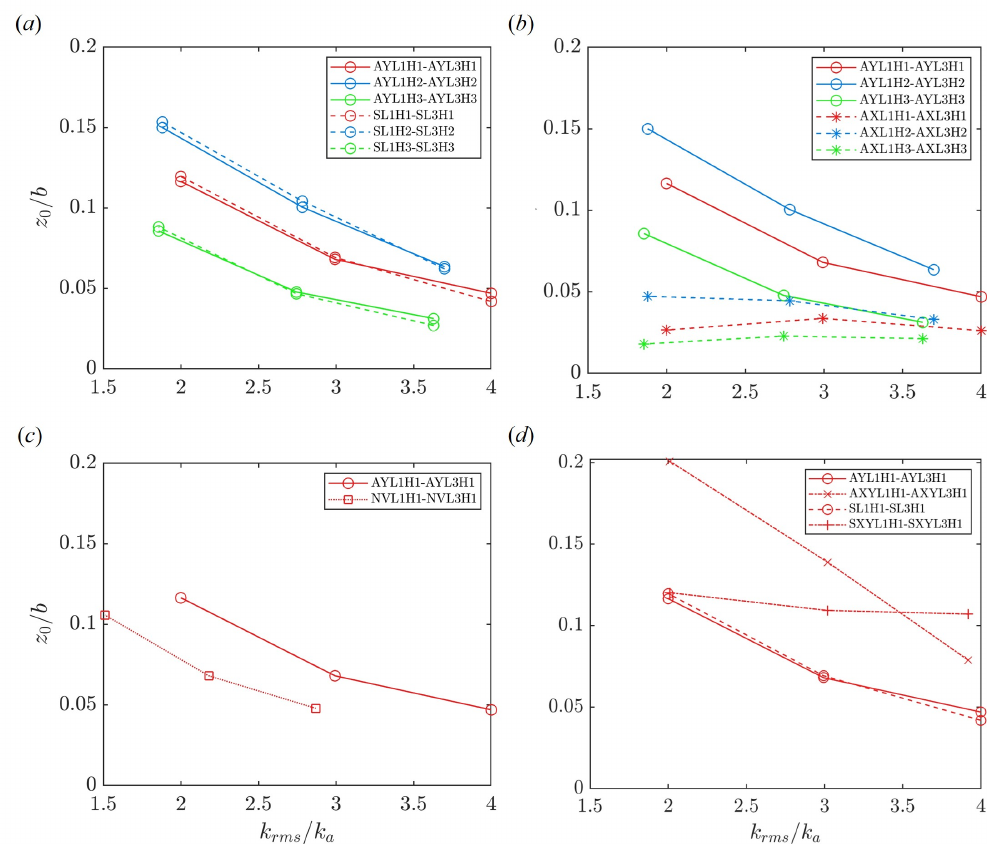}}
  \caption{The effective roughness length to width ratio ($z_0/b$) for the various cases: (a) Spanwise aligned vs Staggered, (b) Spanwise aligned vs Streamwise aligned, (c) Spanwise aligned with and without valleys, and (d) Spanwise aligned/Staggered and their $45^\circ$ rotated equivalents.}
\label{fig:z0_var}
\end{figure}

Figure \ref{fig:z0_var} summarizes the $z_0$ magnitudes for all cases, and the data are consistent with the various mean velocity profiles observed so far. 
Figure \ref{fig:z0_var}(a) compares the spanwise aligned (AY), and staggered (S) cases. 
The $z_0/b$ magnitudes are observed to be similar between the AY and the S cases when comparing cases with similar $\lambda_p$ and $Sk$ magnitudes.
When cases with similar packing densities $\lambda_p$ and orientation but different $Sk$ are compared, for instance, (AYL1H1, AYL1H2 and AYL1H3), the $z_0/b$ increases with $Sk$. 
This holds true for all cases listed in table \ref{tab:DNS results}.
When cases with similar $Sk$ and orientation but different $\lambda_p$ are compared, for instance, (AYL1H1, AYL2H1 and AYL3H1), the $z_0/b$ decreases with decreasing $k_{rms}/k_a$. 
This is observed true for all cases except the AX surfaces, which will be discussed subsequently. 
It should be of interest to note that low $k_{rms}/k_a$ magnitudes actually correspond to high $k_{rms}/b$ and $\lambda_p$. 
The normalisation with $k_a$ causes this as an increment in $\lambda_p$ increases both $k_a$ and $k_{rms}$, but $k_a$ increases at a rate faster than $k_{rms}$.
Figure \ref{fig:z0_var}(b) compares the aligned-spanwise (AY) and aligned-streamwise (AX) cases. 
The effects due to changes in roughness element alignment are evident as the AX surfaces in figure \ref{fig:z0_var}(b) show notably lower $z_0/b$ magnitudes when compared to AY surfaces. 
For instance, even the densely packed surfaces AXL3H1, AXL3H2, and AXL3H3, which have higher solidities, are lower in $z_0/b$ than the coarsely packed surfaces AYL1H1, AYL1H2, and AYL1H3.
Moreover, it can be noticed that $z_0/b$ first increases then decreases with decreasing $k_{rms}/k_{a}$ (or increasing $\lambda_p$) for H1 and H3 cases in AX surfaces.
For higher $\lambda_p$ surfaces AXL2H2 and AXL3H2, the increase in $z_0/b$ is not as significant as it is for AYL2H2 and AYL3H3.
These observations showcase a flow-sheltering effect in AX surfaces. 
Trends with changes in $Sk$ for AX surfaces are similar to that observed in AY ones, with $z_0/b$ increasing with increasing $Sk$.
Figure \ref{fig:z0_var}(c) reports the effect of valleys on $z_0/b$. 
The presence of valleys in the AY cases leads to higher $k_{rms}/k_a$ compared to the NV cases, but the $z_0/b$ magnitudes are very close between the AY and NV cases, with some noticeable difference at low $k_{rms}/k_a$ magnitudes.
Figure \ref{fig:z0_var}(d) presents aligned-spanwise (AY), $45^\circ$ alignment (AXY) and their respective staggered equivalents (S and SXY). 
A significant increase in $z_0/b$ can be observed as alignment is changed to $45^\circ$ when comparing AY and AXY. 
SXY falls in the intermediate, with $z_0/b$ increasing due to a change in alignment when compared with S.
However, this effect is less pronounced as $\lambda_p$ increases or $k_{rms}/k_a$ decreases.

\begin{table}
\begin{center}
\def~{\hphantom{0}}
        \begin{tabular}{lcccccccccccccc}
        Case  & $k_a/b$ & $k_{rms}$/$k_a$ & $Sk$ & $Ku$ & $k_t$/$k_a$ & $ES$ & $Rl_x$/$k_a$ & $Rl_y$/$k_a$ & $d/b$ & $z_0/b$  & ~ & ~ & ~ & ~ \\[3pt] 
        AYL1H1 & 0.063 & 4.00 & 0.00 & 16.0 & 31.98 & 0.125 & 127.9 & 63.95 & 0.182 & 0.047  & ~ & ~ & ~ & ~ \\ 
        AYL2H1 & 0.111 & 2.99 & 0.00 & 9.00 & 17.97 & 0.222 & 53.92 & 26.96 & 0.361 & 0.068  & ~ & ~ & ~ & ~ \\ 
        AYL3H1 & 0.250 & 2.00 & 0.00 & 4.00 & 7.99 & 0.500 & 15.99 & 7.990 & 0.549 & 0.117  & ~ & ~ & ~ & ~ \\ 
        AYL1H2 & 0.075 & 3.70 & 1.01 & 16.4 & 29.45 & 0.138 & 107.0 & 53.49 & 0.296 & 0.064  & ~ & ~ & ~ & ~ \\ 
        AYL2H2 & 0.132 & 2.78 & 0.72 & 9.22 & 16.63 & 0.244 & 45.31 & 22.65 & 0.444 & 0.101  & ~ & ~ & ~ & ~ \\ 
        AYL3H2 & 0.293 & 1.88 & 0.41 & 4.06 & 7.50 & 0.550 & 13.62 & 6.81  & 0.703 & 0.150  & ~ & ~ & ~ & ~ \\ 
        AYL1H3 & 0.062 & 3.63 & -1.23 & 16.7 & 28.90 & 0.113 & 128.3 & 64.16 & 0.149 & 0.031  & ~ & ~ & ~ & ~ \\ 
        AYL2H3 & 0.110 & 2.75 & -0.88 & 9.32 & 16.38 & 0.200 & 54.59 & 27.26 & 0.218 & 0.048  & ~ & ~ & ~ & ~ \\ 
        AYL3H3 & 0.244 & 1.86 & -0.49 & 4.09 & 7.39 & 0.450 & 16.38 & 8.19 & 0.424 & 0.086  & ~ & ~ & ~ & ~ \\ 
        SL1H1 & 0.063 & 4.00 & 0.00 & 16.0 & 31.98 & 0.125 & 127.9 & 127.9 & 0.234 & 0.042  & ~ & ~ & ~ & ~ \\ 
        SL2H1 & 0.111 & 2.99 & 0.00 & 9.00 & 17.97 & 0.222 & 53.92 & 53.92 & 0.293 & 0.069  & ~ & ~ & ~ & ~ \\ 
        SL3H1 & 0.250 & 2.00 & 0.00 & 4.00 & 7.99 & 0.500 & 15.99 & 15.99 & 0.474 & 0.120  & ~ & ~ & ~ & ~ \\ 
        SL1H2 & 0.075 & 3.70 & 1.01 & 16.4 & 29.45 & 0.138 & 107.0 & 107.0 & 0.300 & 0.062  & ~ & ~ & ~ & ~ \\ 
        SL2H2 & 0.132 & 2.78 & 0.72 & 9.22 & 16.63 & 0.244 & 45.31 & 45.31 & 0.391 & 0.104  & ~ & ~ & ~ & ~ \\ 
        SL3H2 & 0.293 & 1.88 & 0.41 & 4.06 & 7.50 & 0.550 & 13.62 & 13.62 & 0.620 & 0.154  & ~ & ~ & ~ & ~ \\ 
        SL1H3 & 0.062 & 3.63 & -1.23 & 16.7 & 28.90 & 0.113 & 128.3 & 128.3 & 0.163 & 0.027  & ~ & ~ & ~ & ~ \\ 
        SL2H3 & 0.110 & 2.75 & -0.88 & 9.32 & 16.38 & 0.200 & 54.59 & 54.59 & 0.183 & 0.047  & ~ & ~ & ~ & ~ \\
        SL3H3 & 0.244 & 1.86 & -0.49 & 4.09 & 7.39 & 0.450 & 16.38 & 16.38 & 0.350 & 0.088  & ~ & ~ & ~ & ~ \\ 
        AXL1H1 & 0.063 & 4.00 & 0.00 & 16.0 & 31.98 & 0.125 & 63.95 & 127.9 & 0.243 & 0.026  & ~ & ~ & ~ & ~ \\ 
        AXL2H1 & 0.111 & 2.99 & 0.00 & 9.00 & 17.97 & 0.222 & 26.96 & 53.92 & 0.287 & 0.034  & ~ & ~ & ~ & ~ \\ 
        AXL3H1 & 0.250 & 2.00 & 0.00 & 4.00 & 7.99 & 0.500 & 7.99 & 15.99 & 0.555 & 0.027  & ~ & ~ & ~ & ~ \\ 
        AXL1H2 & 0.075 & 3.70 & 1.01 & 16.4 & 29.45 & 0.138 & 53.49 & 107.0 & 0.417 & 0.033  & ~ & ~ & ~ & ~ \\ 
        AXL2H2 & 0.132 & 2.78 & 0.72 & 9.22 & 16.63 & 0.244 & 22.65 & 45.31 & 0.473 & 0.045  & ~ & ~ & ~ & ~ \\ 
        AXL3H2 & 0.293 & 1.88 & 0.41 & 4.06 & 7.50 & 0.550 & 6.81 & 13.62 & 0.789 & 0.047  & ~ & ~ & ~ & ~ \\ 
        AXL1H3 & 0.062 & 3.63 & -1.23 & 16.7 & 28.90 & 0.113 & 64.16 & 128.3 & 0.076 & 0.021  & ~ & ~ & ~ & ~ \\ 
        AXL2H3 & 0.110 & 2.75 & -0.88 & 9.32 & 16.38 & 0.200 & 27.26 & 54.59 & 0.272 & 0.023  & ~ & ~ & ~ & ~ \\ 
        AXL3H3 & 0.244 & 1.86 & -0.49 & 4.09 & 7.39 & 0.450 & 8.19 & 16.38 & 0.396 & 0.018  & ~ & ~ & ~ & ~ \\  
        AXYL1H1 & 0.065 & 3.92 & 0.00 & 15.5 & 30.90 & 0.176 & 180.9 & 180.9 & 0.080 & 0.079  & ~ & ~ & ~ & ~ \\ 
        AXYL2H1 & 0.110 & 3.02 & 0.00 & 9.08 & 18.21 & 0.307 & 76.26 & 76.26 & 0.005 & 0.139  & ~ & ~ & ~ & ~ \\ 
        AXYL3H1 & 0.247 & 2.01 & 0.00 & 4.04 & 8.09 & 0.688 & 22.61 & 22.61 & 0.070 & 0.201  & ~ & ~ & ~ & ~ \\ 
        SXYL1H1 & 0.065 & 3.92 & 0.00 & 15.48 & 30.90 & 0.176 & 90.45 & 90.45 & -0.424 & 0.107  & ~ & ~ & ~ & ~ \\ 
        SXYL2H1 & 0.110 & 3.02 & 0.00 & 9.08 & 18.21 & 0.307 & 38.13 & 38.13 & 0.128 & 0.109  & ~ & ~ & ~ & ~ \\ 
        SXYL3H1 & 0.247 & 2.01 & 0.00 & 4.04 & 8.09 & 0.688 & 11.31 & 11.31 & 0.535 & 0.120  & ~ & ~ & ~ & ~ \\
        NVL1H1 & 0.060 & 2.87 & 5.88 & 30.0 & 16.56 & 0.063 & 132.5 & 66.27 & 0.132 & 0.048  & ~ & ~ & ~ & ~ \\ 
        NVL2H1 & 0.105 & 2.18 & 3.88 & 16.1 & 9.55 & 0.111 & 57.18 & 28.59 & 0.350 & 0.068  & ~ & ~ & ~ & ~ \\ 
        NVL3H1 & 0.219 & 1.51 & 2.27 & 6.14 & 4.57 & 0.250 & 18.29 & 9.14 & 0.636 & 0.106  & ~ & ~ & ~ & ~ \\
        
    \end{tabular}
    \caption{Roughness' hydrodynamic properties.
    The geometric parameters listed include: average and root-mean-square roughness height $k_a$ and $k_{rms}$, skewness $Sk$, kurtosis $Ku$, maximum peak-to-trough height $k_t$, effective slope(along $x$ direction) $ES$, streamwise and spanwise correlation lengths $Rl_{x}$ and $Rl_y$ (defined in \ref{sub:tp}). $d$ denotes zero-plane displacement height and $z_0$ the effective roughness height.}
  \label{tab:DNS results}
  \end{center}
\end{table}

\subsection{Reynolds and dispersive stresses}

We present the Reynolds stress and the dispersive stress results.
These results are usually not relevant to roughness modelling but are fundamental aspects of the flow, and we present them here for completeness.

\begin{figure}
  \centerline{\includegraphics[width=0.9\linewidth]{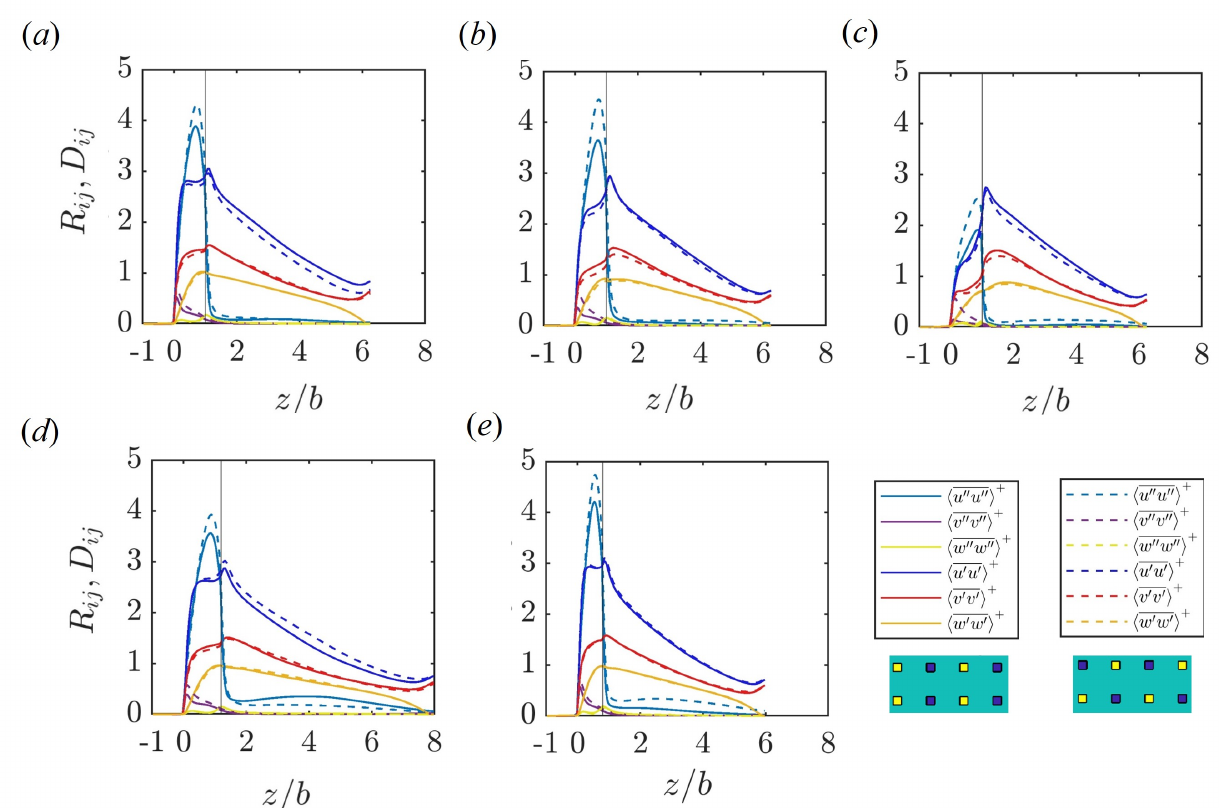}}
  \caption{Reynolds $\langle \overline{u_i'u_j'}\rangle^+$ and dispersive $\langle \overline{u_i''u_j''}\rangle^+$ stress comparisons for spanwise aligned and staggered cases. The solid and dashed lines depict aligned and staggered arrangements respectively. (\textit{a}) AYL1H1 and SL1H1, (\textit{b}) AYL2H1 and SL2H1, (\textit{c}) AYL3H1 and SL3H1, (\textit{d}) AYL1H2 and SL1H2, and (\textit{e}) AYL1H3 and SL1H3. The vertical line indicates peak roughness height.}
\label{fig:Re-Disp_stressAS}
\end{figure}

Figure \ref{fig:Re-Disp_stressAS} shows the normal components of the horizontally averaged Reynolds and dispersive stresses for the five spanwise aligned (AYL1H1, AYL2H1, AYL3H1, AYL1H2, AYL1H3) and staggered (SL1H1, SL2H1, SL3H1, SL1H2, SL1H3) cases.
At first glance, all stresses are negligible in the valleys below the surface at $z=0$, and the spanwise and wall-normal components of the dispersive stresses are small throughout the channel.
Second, it can be observed that the corresponding streamwise components of both stresses ($uu$) are higher than the respective spanwise ($vv$) and wall-normal ($ww$) components for all cases.
Third, the maximum value for dispersive stress lies within the roughness-occupied layer whereas for the Reynolds stress, the maxima lie in the neighbourhood of the roughness peak.
Moreover, the dispersive stress spike for staggered cases is slightly higher than the aligned ones.
This spike also decreases with packing density for this set of cases. 
This can be attributed to the reduced inhomogeneity in the mean velocity field as the spacing between cubes reduces.
Comparing zero-skewed surface (figures \ref{fig:Re-Disp_stressAS} \textit{a}) with the corresponding positively skewed (figures \ref{fig:Re-Disp_stressAS} \textit{d}) and negatively skewed ones (figures \ref{fig:Re-Disp_stressAS} \textit{e}), it can be seen that dispersive stress spike gets narrower and increases with decreasing $Sk$.
The narrowing is an indirect consequence of the decrease in roughness height ($h_1/b$) as $Sk$ decreases.
The Reynolds stresses, on the other hand, are similar irrespective of packing density or skewness. 

\begin{figure}
  \centerline{\includegraphics[width=0.9\linewidth]{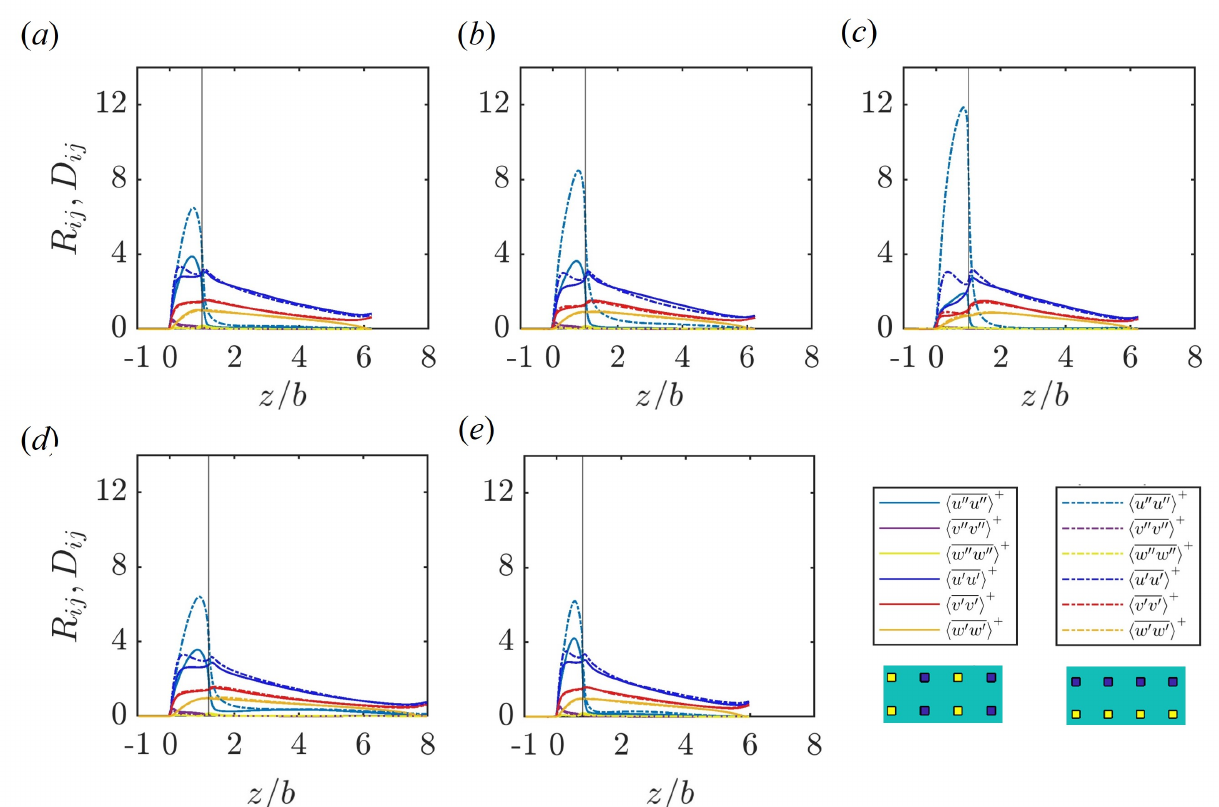}}
  \caption{Reynolds $\langle \overline{u_i'u_j'}\rangle^+$ and dispersive $\langle \overline{u_i''u_j''}\rangle^+$ stress comparisons for spanwise aligned and streamwise aligned cases. The solid and dashed lines depict spanwise-aligned and streamwise-aligned arrangements respectively.  (\textit{a}) AYL1H1 and AXL1H1, (\textit{b}) AYL2H1 and AXL2H1, (\textit{c}) AYL3H1 and AXL3H1, (\textit{d}) AYL1H2 and AXL1H2, and (\textit{e}) AYL1H3 and AXL1H3. The vertical line indicates peak roughness height.}
\label{fig:Re-Disp_stressAXAY}
\end{figure}

\begin{figure}
  \centerline{\includegraphics[width=0.95\linewidth]{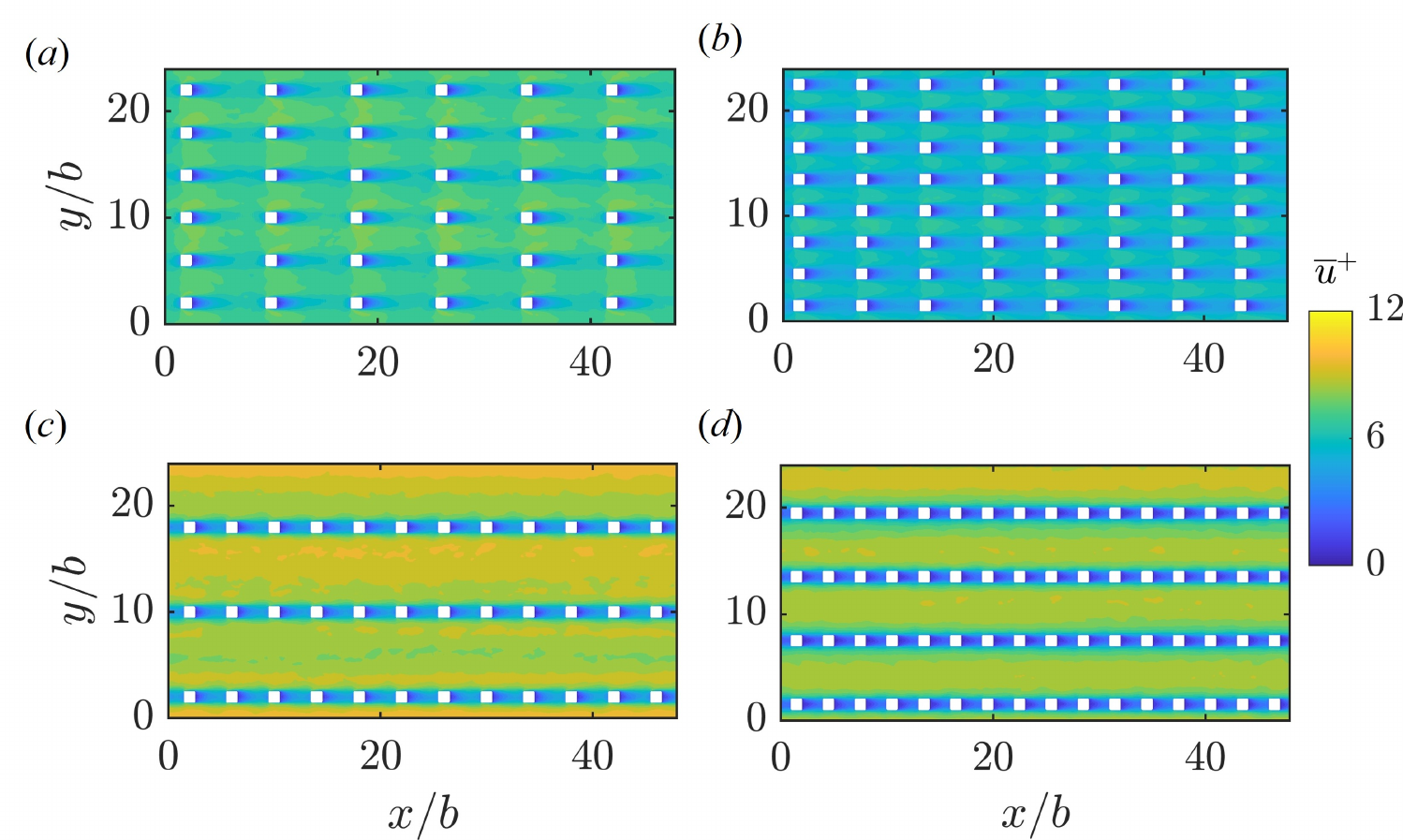}}
  \caption{Streamwise mean velocity contours at $z/b \approx 1.0$ for cases: (\textit{a}) AYL1H1, (\textit{b}) AYL2H1, (\textit{c}) AXL1H1 and (\textit{d}) AXL2H1.}
\label{fig:mean_vel_contour}
\end{figure}

Similarly, plots comparing the stresses for spanwise aligned (AY) and streamwise aligned (AX) surfaces are shown in figure \ref{fig:Re-Disp_stressAXAY}.
The higher magnitudes of dispersive stress for AX surfaces are due to the presence of high momentum pathways between two rows of cubes and the low momentum pathways over cube locations. 
Unlike figure \ref{fig:Re-Disp_stressAS}, the dispersive stress spike here increases with packing density for the AX surfaces.
To further probe these observations, the streamwise mean velocity contours for AYL1H1, AYL2H1, AXL1H1 and AXL2H1 near the roughness peaks are shown in figures \ref{fig:mean_vel_contour} (\textit{a}), (\textit{b}), (\textit{c}) and (\textit{d}) respectively. 
Indeed, we see more inhomogeneity in the mean flow in \ref{fig:mean_vel_contour} (\textit{c}) and (\textit{d})  due to the overlapping wake regions and the high momentum pathways formed in between the cubes.
With respect to skewness, the increasing trend of dispersive stress with decreasing skewness for AX surfaces is similar to that observed in AY surfaces.

\begin{figure}
  \centerline{\includegraphics[width=0.95\linewidth]{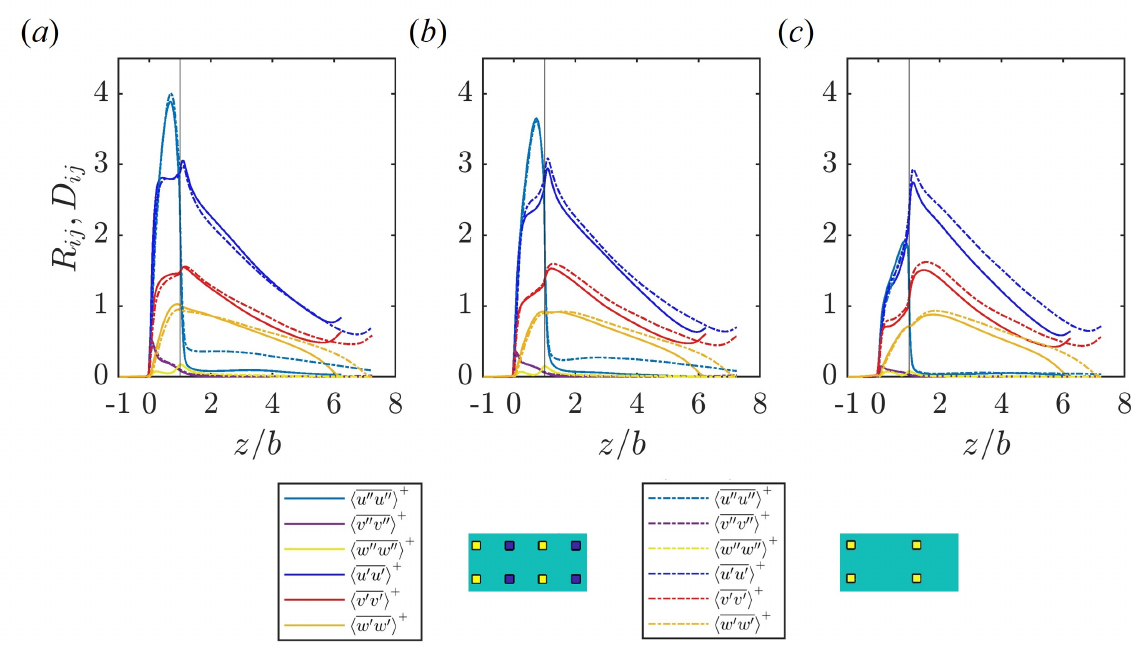}}
  \caption{Reynolds $\langle \overline{u_i'u_j'}\rangle^+$ and dispersive $\langle \overline{u_i''u_j''}\rangle^+$ stress comparisons for spanwise-aligned cases with and without valleys: (\textit{a}) AYL1H1/NVL1H1, (\textit{b}) AYL2H1/NVL2H1 and (\textit{c}) AYL3H1/NVL3H1. The solid and dashed lines depict spanwise-aligned and no-valleys cases respectively. The vertical line indicates peak roughness height.}
\label{fig:Eff_valleys_RD}
\end{figure}

The normal stress comparisons in figure \ref{fig:Eff_valleys_RD} show slight variations when valleys are removed from AYL1H1, AYL2H1 and AYL3H1.
Note that the NV surfaces have very different roughness statistics in $k_{rms}/k_a$, $Sk$, $Ku$ and $ES$ but very similar Reynolds and dispersive stresses (figure \ref{fig:Eff_valleys_RD}), mean velocity profiles (figure \ref{fig:Eff_valley}) and $z_0/b$ (figure \ref{fig:z0_var}(\textit{c})) with respect to the AY and S counterparts.

\begin{figure}
  \centerline{\includegraphics[width=\linewidth]{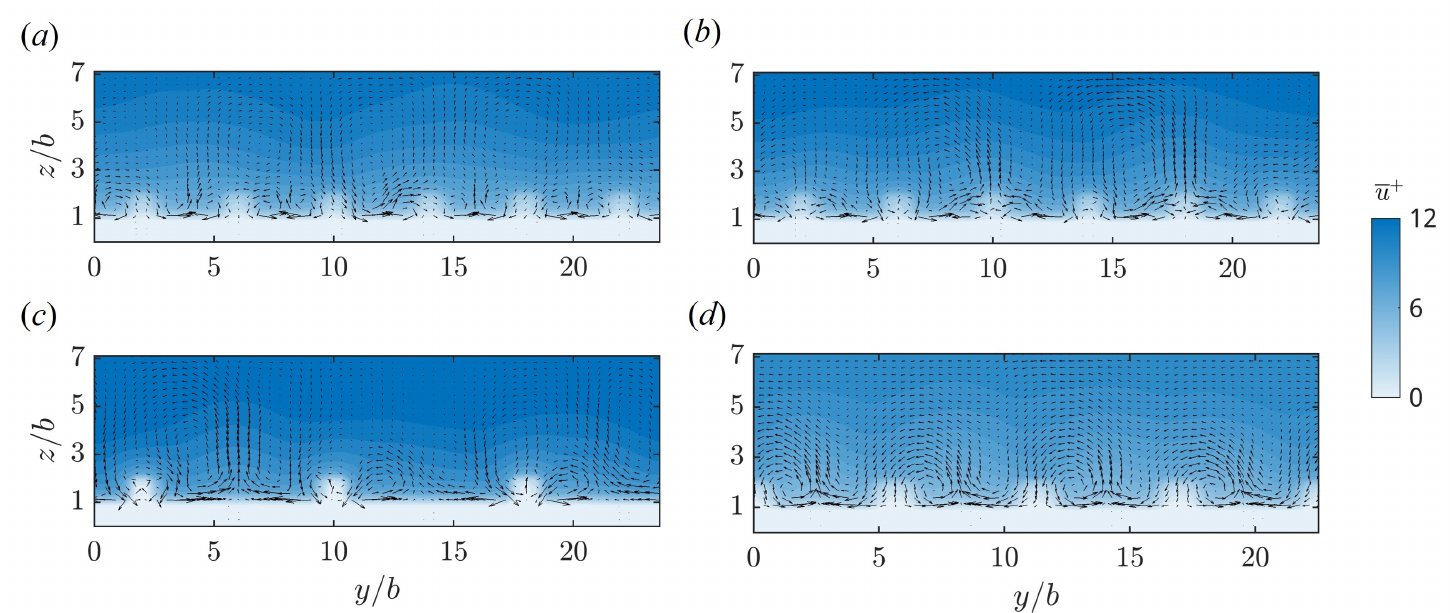}}
  \caption{Mean streamwise averaged constant $x$-plane showcasing $\overline{u}^+$ contours with secondary flows $\overline{v}^+$ and $\overline{w}^+$ (denoted by vectors) within the domain in: (\textit{a}) AYL1H1, (\textit{b}) SL1H1, (\textit{c}) AXYL1H1 and (\textit{d}) SXYL1H1.}
\label{fig:Sec_flow2}
\end{figure}

In figures \ref{fig:Re-Disp_stressAS}, \ref{fig:Re-Disp_stressAXAY} and \ref{fig:Eff_valleys_RD}, it can be observed that the streamwise dispersive component is still measurable near the half-channel height.
These effects hint at the presence of secondary flows within these domains.
Figure \ref{fig:Sec_flow2} shows the mean streamwise averaged contours of $\overline{u}^+$ with secondary flow $\overline{v}^+$ and $\overline{w}^+$ for a few cases. 
One can observe small but non-zero secondary flow near the half-height and noticeable counter-rotating vortical structures near the roughness peaks.
In most cases, the secondary structures can be seen to penetrate beyond the roughness sublayer, which is in alignment with other studies such as \cite{vanderwel_ganapathisubramani_2015}.

\section{Rough-wall Modelling}\label{sec:4}

It should be clear from the results presented in \S \ref{sec:3} that the roughness regime studied here, while previously unexplored, has no unexpected behaviors.
In this section, we discuss the implications the data has on rough-wall modelling.
In addition to the DNS data here, we also make use of the data in the roughness database (\url{https://roughnessdatabase.org/}).
While it will be clear later in this section, the task of rough-wall modelling is the task of identifying roughness statistics, such as $k_{\rm rms}$ and $Sk$, that distinguish all rough surfaces in the calibration set.
As it is highly unlikely that a finite set of roughness statistics can distinguish all rough surfaces, it is improbable that a universal rough-wall model that perfectly captures all rough surfaces can be formulated.
Hence, instead of a universal rough wall model, it is more effective to seek rough wall models with clearly defined applicable ranges.

\subsection{The task of roughness modelling}
\label{sub:tp}

In this subsection, we try to answer the following question: what is the nature of roughness modeling?
We begin by comparing the $z_o$ magnitudes in Table \ref{tab:DNS results} to the estimates provided by the existing rough-wall correlations.
Here, the correlations in \cite{Forooghi2017TowardCorrelation} and \cite{flack2020skin} are taken as examples.
The two correlations are given in \eqref{eq:ks forooghi} and \eqref{eq:ks flack} respectively:
\begin{equation}\label{eq:ks forooghi}
  k_s/k_t = 1.07\cdot(1-e^{-3.5 \cdot ES})(0.67Sk^2 + 0.93Sk +1.3).
\end{equation}
\begin{equation}\label{eq:ks flack}
  k_s = \left\{
    \begin{array}{lll}
      2.11 k_{rms}, & Sk = 0 \\[2pt]
      2.48 k_{rms}(1+Sk)^{2.24}, & Sk > 0 \\
      2.73 k_{rms}(2+Sk)^{-0.45} & Sk < 0  
  \end{array} \right.
\end{equation}
where roughness statistics $k_{rms}$, $Sk$, $k_t$, and $ES$ are invoked as inputs to model $k_s$.
We note that the effective roughness height ($z_0$) in Table \ref{tab:DNS results} is directly linked to $k_s$ by the following expression,
\begin{equation}
    k_s = z_0 e^{(\kappa A)}
\end{equation}
where the von-K\'arm\'an constant $\kappa = 0.4$  and $A = 8.5$.

\begin{figure}\centerline{\includegraphics[width=0.9\linewidth]{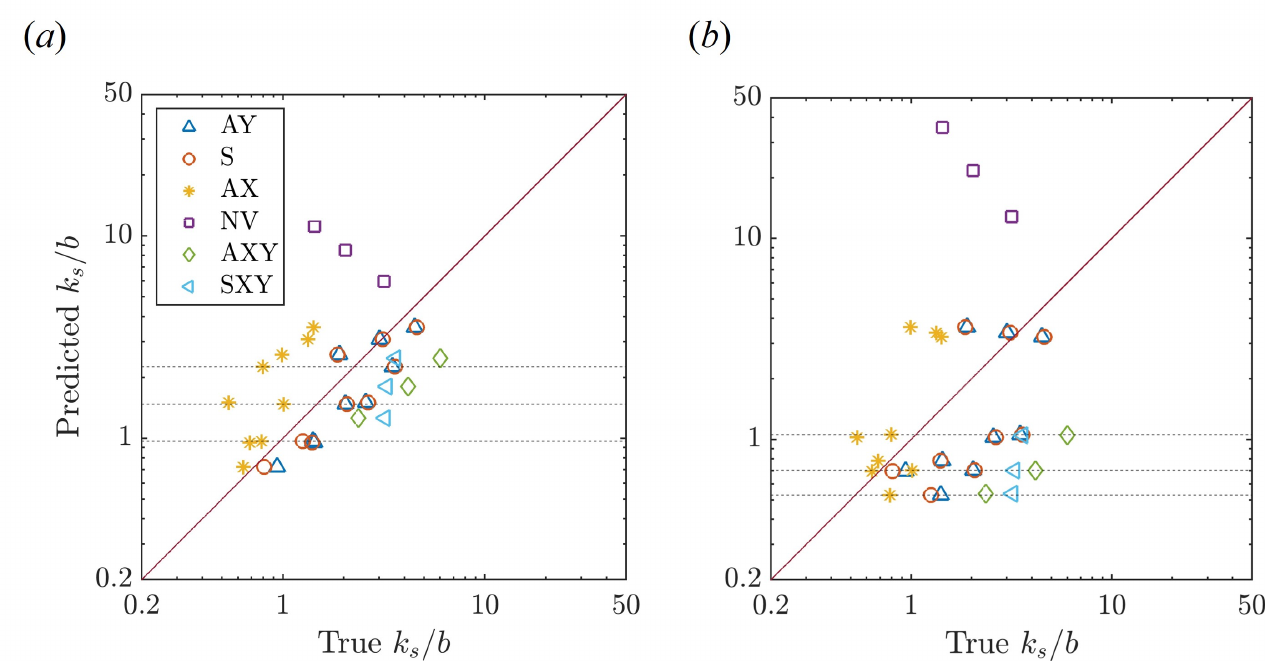}}
  \caption{Existing Correlations by (a) \cite{Forooghi2017TowardCorrelation}  and (b) \cite{flack2020skin}, applied to surface parameters obtained from the current study.
  The plots show the predicted equivalent sandgrain roughness height $k_s$ normalized by cube width $b$. }
\label{fig:Existing Correlations}
\end{figure}

Figure \ref{fig:Existing Correlations} shows the predicted $k_s$ for the rough surfaces in this study using the aforementioned correlations.
We see that the surfaces in the present study do not fit into these existing correlations.
Further, we see from figure \ref{fig:Existing Correlations}(a) that surfaces with different arrangements (AX, AY, S), although have very different measured $k_s$, produce the same estimates according to \eqref{eq:ks forooghi} as they have the same $k_t$, $Sk$ and $ES$ (as indicated by the dashed lines). 
Similarly, in figure \ref{fig:Existing Correlations}(b), the surfaces AXY and SXY, although have very different measured $k_s$, join the other arrangements with identical $k_{rms}$ and $Sk$, resulting in identical $k_s$ predictions according to \eqref{eq:ks flack}.
These observations indicate the inherent lack of roughness statistics that distinguish roughness surfaces in these correlations. 
Although the discussion here is limited to the correlations in \cite{Forooghi2017TowardCorrelation} and \cite{flack2020skin}, the same inadequacies are prevalent in other empirical correlations as well, as observed by \cite{yang2023search}.

\begin{figure}
  \centerline{\includegraphics[width=0.85\linewidth]{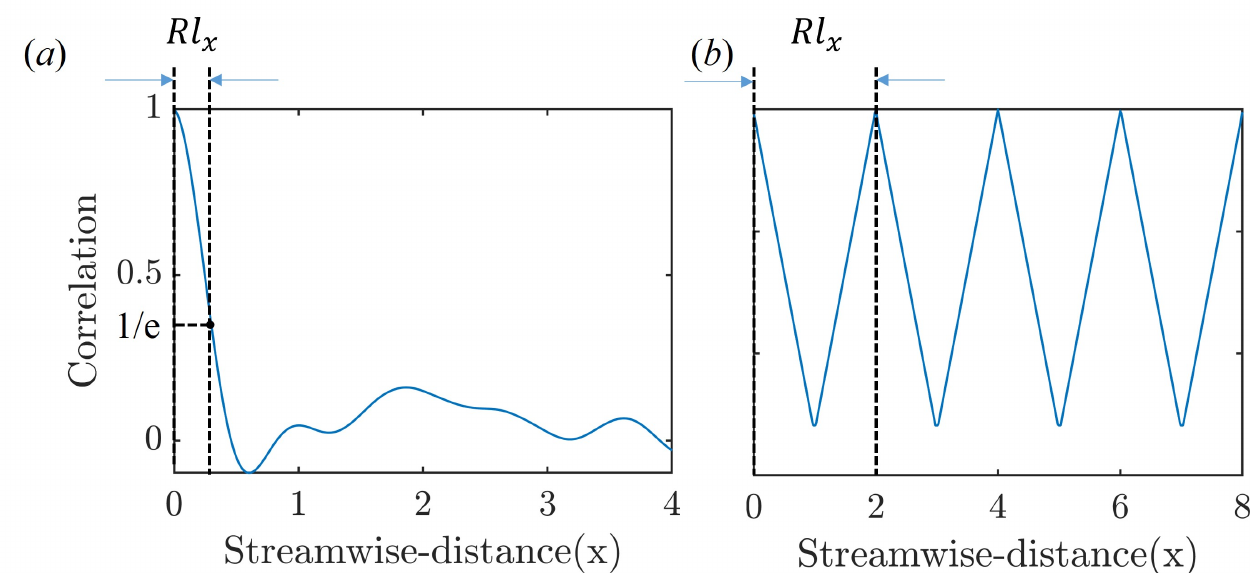}}
  \caption{An illustration of the definition of correlation length for (a) irregular surface and (b) regular surface.}
\label{fig:corr_length_def}
\end{figure}

A remedy, as advocated and explained in \cite{Chung2021PredictingSurfaces} and other references cited therein, is to expand the list of inputs and include other roughness statistics.
Here, statistics that help to distinguish different arrangements of elements are needed. 
Among other statistics, two-point correlations are the simplest ones for this purpose.
The conventional definition of correlation length measures the distance at which the auto-correlation,
\begin{equation}\label{Cx}
C_x(\Delta x) = \frac{1}{L_x L_y k_{rms}^2}\int_{L_y} \int_{L_x} (k(x, y) - \overline{k})(k(x + \Delta x, y) - \overline{k}) dx dy,
\end{equation}
\begin{equation}\label{Cy}
C_y(\Delta y) = \frac{1}{L_x L_y k_{rms}^2}\int_{L_y} \int_{L_x} (k(x, y) - \overline{k})(k(x, y + \Delta y) - \overline{k}) dx dy
\end{equation}
drops to a certain value, typically $1/e$. 
Here, $k(x,y)$ contains the height information for the rough surface at location $(x, y)$.
In other words, the correlation lengths $Rl_x$ and $Rl_y$ are such that 
\begin{equation}\label{rlx}
C_x(Rl_x) = 1/e,
\end{equation}
\begin{equation}\label{rly}
C_y(Rl_y) = 1/e,
\end{equation}
The above definitions are suitable for irregular surfaces, but when applied to regular, e.g., cuboidal roughness, thus-defined correlations simply give the roughness element width.
Intuitively, we need a distance measure that is representative of the inter-repeating-tile length.
This can be made possible by choosing the peak-to-peak distance in the correlation plot.
Figure \ref{fig:corr_length_def} demonstrates these definitions, with an irregular surface A0020 from \cite{Forooghi2017TowardCorrelation}, and a regular surface R25 from \cite{xu2021flow}, as examples.
Extending them to both streamwise and spanwise directions, one can distinguish between staggered, aligned as well as surfaces with different arrangements.

\begin{figure}
  \centerline{\includegraphics[width=0.9\linewidth]{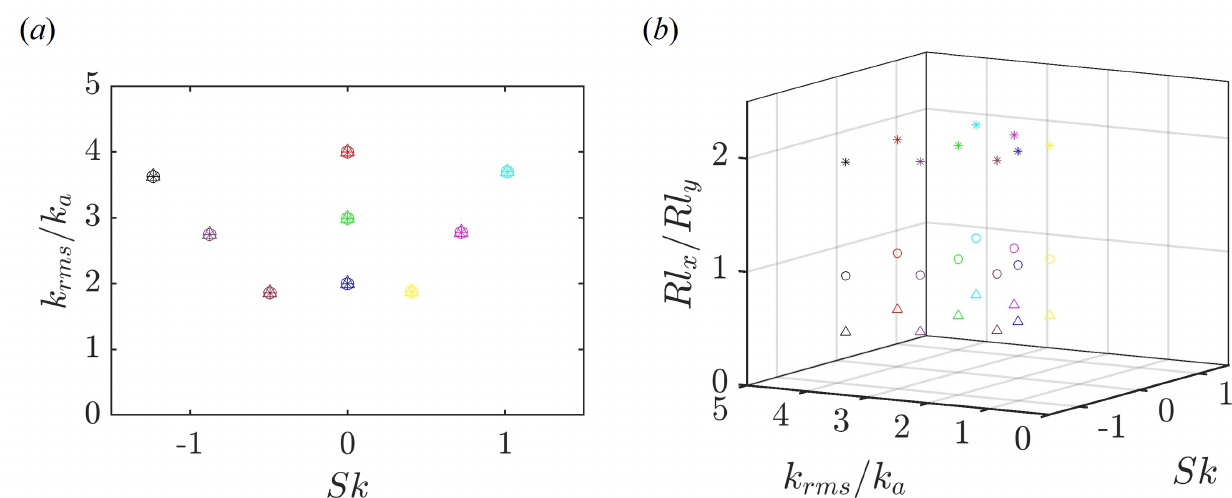}}
  \caption{Rough surfaces in two different parameter spaces. (a) $k_{rms}/k_a$ and $Sk$, and (b) $k_{rms}/k_a$, $Sk$ and $Rl_x/Rl_y$.}
\label{fig:3pcorr_story}
\end{figure}

Having defined the correlation lengths, it is now possible to distinctly distinguish between the rough surfaces involved in this study.
We demonstrate this in Figure \ref{fig:3pcorr_story}, where we look at 27 surfaces from our DNS study, namely AYL$i$H$j$, SL$i$H$j$ and AXL$i$H$j$, with i, j = 1,2,3.
The number of input roughness statistics are progressively increased so that figures \ref{fig:3pcorr_story}(\textit{a}) contains $k_{rms}/k_a$ and $Sk$ and \ref{fig:3pcorr_story}(\textit{b}) contains $k_{rms}/k_a$, $Sk$ and $Rl_x/Rl_y$. 
We observe that the data points overlap in figure \ref{fig:3pcorr_story}(\textit{a}) as AY/AX/SL$i$H$j$ occupy the same location in the space, but in figure \ref{fig:3pcorr_story}(\textit{b}), the rough surfaces do not overlap.
Now, if one were to build a rough-wall model using only $Sk$ and $k_{rms}$ as its inputs, one would be seeking different outputs from the same input, which would inevitably result in errors, as observed in figure \ref{fig:Existing Correlations}.
However, if $Sk$, $k_{rms}$, and $Rl_x/Rl_y$ are invoked as inputs, one would be seeking for a single-valued function that maps from the input roughness statistics space to the output $k_s$ space.
Following this line of thinking, we argue that the task of roughness modelling can be reduced to identifying a set of roughness statistics that distinguishes the rough surfaces in question.

This conclusion has many implications for rough-wall modelling.
Firstly, considering that there is not a finite set of statistics that allows unique identification of all rough surfaces, i.e., 2D functions $k(x,y)$, constructing a universal roughness-statistics-based rough-wall model could be a very difficult problem to solve.
Consequently, all roughness-statistics-based rough-wall models are bound to have predictive power for surfaces that are closely registered around its calibration dataset only.
Secondly, roughness modelling must consider the calibration dataset, and a universal list of roughness statistics for all rough-wall models is highly unlikely.
One can argue that certain roughness types are more important than others and therefore certain roughness statistics are more important than others, such judgments will have to be subjective.

\subsection{Selecting roughness statistics}

Considering that all rough-wall models will have their applicable range, selecting inputs to the rough-wall models is a highly relevant issue.
In this subsection, we discuss how one might go about selecting the roughness statistics that distinguish the rough surfaces in a given group of rough surfaces.

\begin{figure}  \centerline{\includegraphics[width=0.9\linewidth]{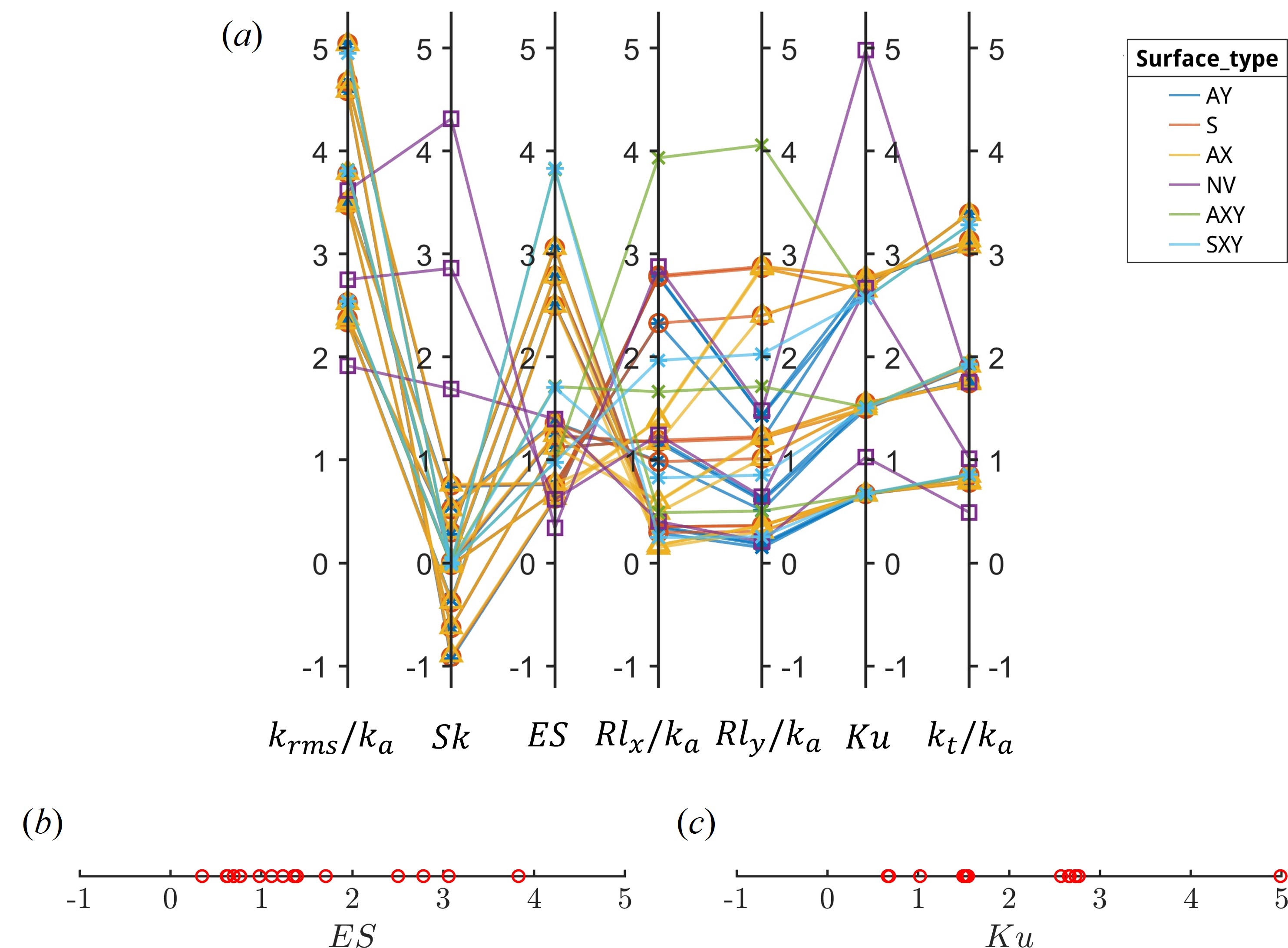}}
  \caption{High-dimensional parallel plots of the surface parameter space for (a) 36 surfaces in the current study (b) $ES$ parameter space and (c) $Ku$ parameter space. The values shown are scaled by the standard deviation. }
\label{fig:ParallelPlot}
\end{figure}

Without loss of generality, we limit the discussion to the 7 roughness statistics in Table \ref{tab:DNS results}.
Figure \ref{fig:ParallelPlot} is a representation of the 36 rough surfaces in the 7-dimensional roughness statistics space represented by $k_{rms}/k_a$, $Sk$, $ES$, $Rl_x/k_a$, $Rl_y/k_a$, $Ku$, and $k_t/k_a$.
In the figure, every roughness statistic corresponds to a vertical axis, and every rough surface corresponds to a line in the figure.
We will make use of this plot to select the roughness statistics that are most relevant to the modeling of the rough surfaces in this study.
We can, of course, use all roughness statistics.
The objective, however, is to identify as few roughness statistics as possible such that the identified roughness statistics distinguish the rough surfaces in question.
It is also worth noting that, for random roughness, $k_t$ might not be a suitable parameter to represent the surface as it tends to increase as the sampling size gets larger instead of converging.

\begin{figure}
  \centerline{\includegraphics[width=0.9\linewidth]{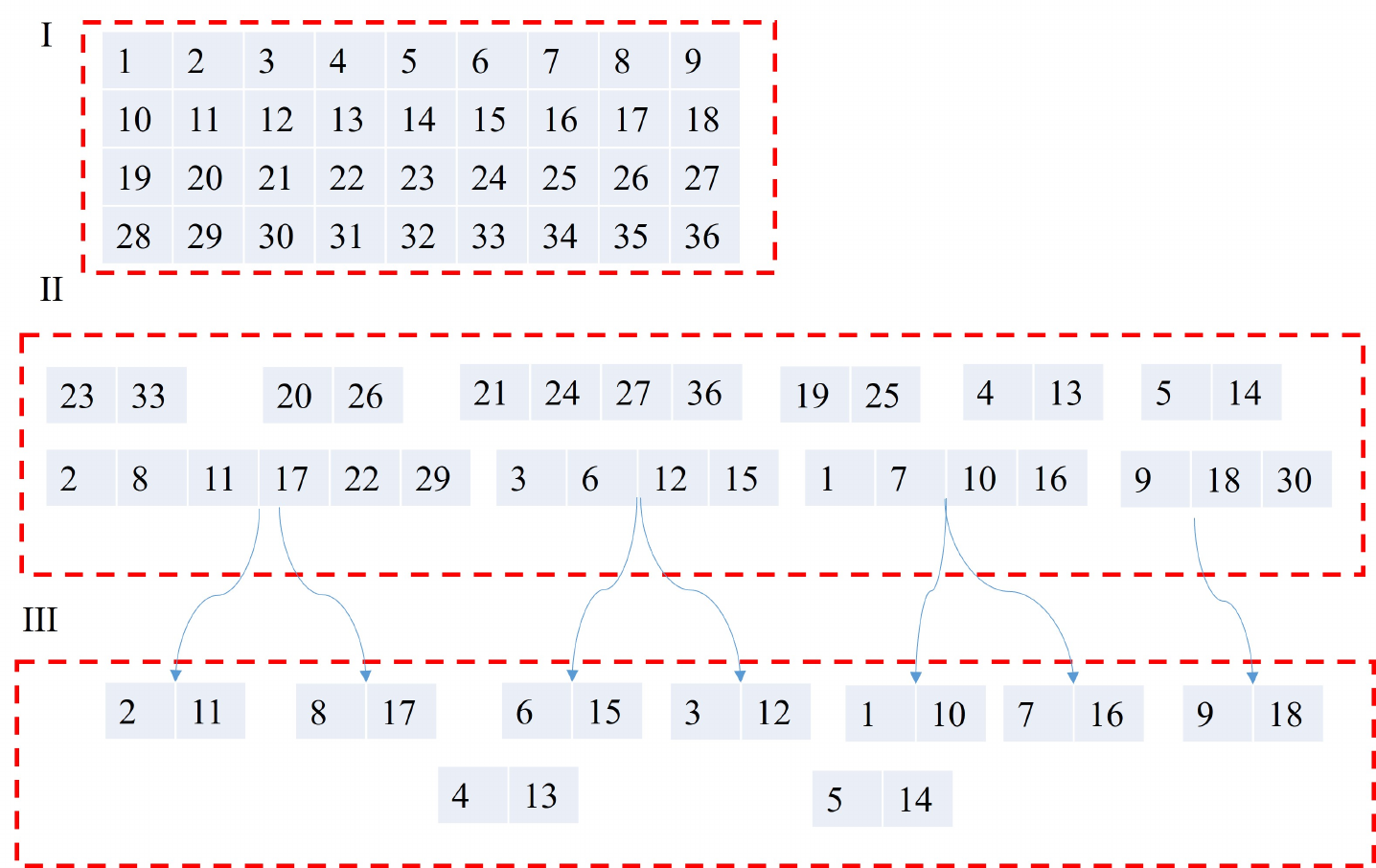}}
  \caption{Feature selection procedure illustration using rough surface indices. Each number here is representative of a rough surface. I denotes the initial stage involving all surfaces in the group, II denotes the next stage after one feature $Rl_x/k_a$ was selected, and III denotes the third stage when $ES$ is selected.}
\label{fig:FeatureSelection_}
\end{figure}

First, we make the following observation:
For a roughness statistic $s$ which represents a vertical axis in figure \ref{fig:ParallelPlot}, rough surfaces that do not overlap in the axis can readily be distinguished by that statistic.
For example, consider two surfaces AYL1H1 and SL1H1.
All features excluding $Rl_y/k_a$ would remain the same for both these surfaces and hence this parameter would serve as the distinguishing feature.
However, depending on how accurately one could measure the features, the features need not necessarily be identical for them to be overlapping. 
We may mathematically define ``overlap'' as $\left|s_i-s_j\right|<c {\rm STD}_s$.
That is, if $s_i$ and $s_j$ are close, they are considered indistinguishable or overlap.
Here, $s_i$ is the roughness statistic of the $i$th rough surface, STD$_s$ is the standard deviation of statistic $s$ in the rough surfaces, and $c$ is a constant. 
Again, the choice of $c$ will depend on how well one can measure the roughness statistic in question: if one can measure the statistic with a precision of $0.01$STD$_s$, $c$ can take the values of 0.02, i.e., +0.01-(-0.01).
Here, we set $c=0.1$, which should be a very conservative choice.
Following this line of thinking, a roughness statistic that distinguishes more rough surfaces is more relevant to the modelling of the rough surfaces in question.
Take the statistics and the rough surfaces in figure \ref{fig:ParallelPlot}(a) as our illustrative example.
Kurtosis is not an extremely relevant roughness statistic for the modelling of these rough surfaces as it only distinguishes 5 groups of rough surfaces (see figure \ref{fig:ParallelPlot}(c)).
On the other hand, from figure \ref{fig:ParallelPlot}(b) it can be observed that ES is a more relevant roughness statistic as it distinguishes around 12 groups of rough surfaces.
In other words, the more the spread in the feature space, higher would be its differentiating ability. 
Note that surfaces indistinguishable by a parameter may be indistinguishable in terms of drag. 
In such cases, the model would remain accurate even if it does not distinguish the two surfaces.

Following the discussion above, the process of selecting roughness statistics that distinguish rough surfaces in a given group of rough surfaces can follow a recursive greedy algorithm. 
To illustrate the algorithm, we take our current study as an example.
For this group of rough surfaces, we have 36 surfaces.
We will represent each rough surface by a number, and the 36 rough surfaces are numbered from 1 to 36.
First, we will split each feature into bins of size 0.1 STD, and the number of surfaces present in each bin is counted.
The feature with the most number of non-zero bins is the most distinguishing feature.
For the 36 rough surfaces, $Rl_x$ is found to be that feature.
If a surface is the only surface in one of the $Rl_x$ bins, that surface is readily distinguishable by $Rl_x$.
However, if there are a couple of rough surfaces in one $Rl_x$ bin, these surfaces, which we call a cluster, cannot be distinguished by $Rl_x$, and we must invoke another roughness statistic to distinguish these surfaces. 
From figure \ref{fig:FeatureSelection_}, it can be seen that there are 10 such clusters, with 1 cluster containing 6 surfaces, 3 clusters containing 4 surfaces each, 1 cluster containing 3 surfaces, and 5 clusters containing 2 surfaces each.
The purpose of the next stage is to find the next feature that best distinguishes these surfaces.
A similar procedure is repeated for this stage for each cluster, and $ES$ is identified as the next most distinguishing feature.
The process continues until either no clusters emerge or all features have been selected.
In this case, this stopping criterion is attained after the third roughness statistics, $Rl_y$, is selected. 
Figure \ref{fig:FeatureSelection_} illustrates the procedure. 

We can apply this recursive greedy algorithm to other groups of surfaces.
Table \ref{tab:Regression results} shows some of these examples, where this feature selection procedure is applied to the rough surfaces from \cite{AghaeiJouybari2021Data-drivenFlows}, \cite{womack2022turbulent}, \cite{xu2021flow} and the current study.
Although we do not consider all surface geometries and other roughness parameters such as spanwise effective slope $ES_z$ and surface porosity $P_0$ mentioned in \cite{AghaeiJouybari2021Data-drivenFlows}, it is interesting to note that we obtain one variable as the distinguishing feature within each of the three pairs ($E_x$, $E_z$), ($P_0$, $Sk$) and ($k_{rms}$, $Ku$) that contain strong correlations in their work, and the algorithm does not pick up two strongly correlated features.
In the 6 rough surfaces in \cite{xu2021flow}, where the primary varying parameter is the planar packing density of the cubes, just one parameter, $k_{rms}/k_a$ in this case, seems to be sufficient.
An interesting observation is that $k_{rms}/k_a$ is not always a distinguishing roughness statistic.
For the surfaces with multiple roughness arrangements, such as that in our study, the dimensionless two-point correlation lengths $Rl_x/k_a$ and $Rl_y/k_a$ emerge as having better distinguishing capability.

\begin{table}
    \begin{center}
        \begin{tabular}{lcc}
        Group  & $n_{\text{surfaces}}$ & Selected features \\
        
        %
        \hline
        Jouybari et. al. & 25 & $Sk, ES, k_{rms}/k_a $ \\
        \hline
        Womack et. al & 16 & $Sk, k_{rms}/k_a,  Rl_x/k_a, ES$  \\
        \hline
        Xu et. al. & 6 & $k_{rms}/k_a$ \\
        \hline
        Current study & 36 & $Rl_x/k_a, Rl_y/k_a, ES $ \\
        
        \end{tabular}
    \caption{Features selected for different groups.}
  \label{tab:Regression results}
  \end{center}
\end{table}

\begin{figure}
  \centerline{\includegraphics[width=1.0\linewidth]{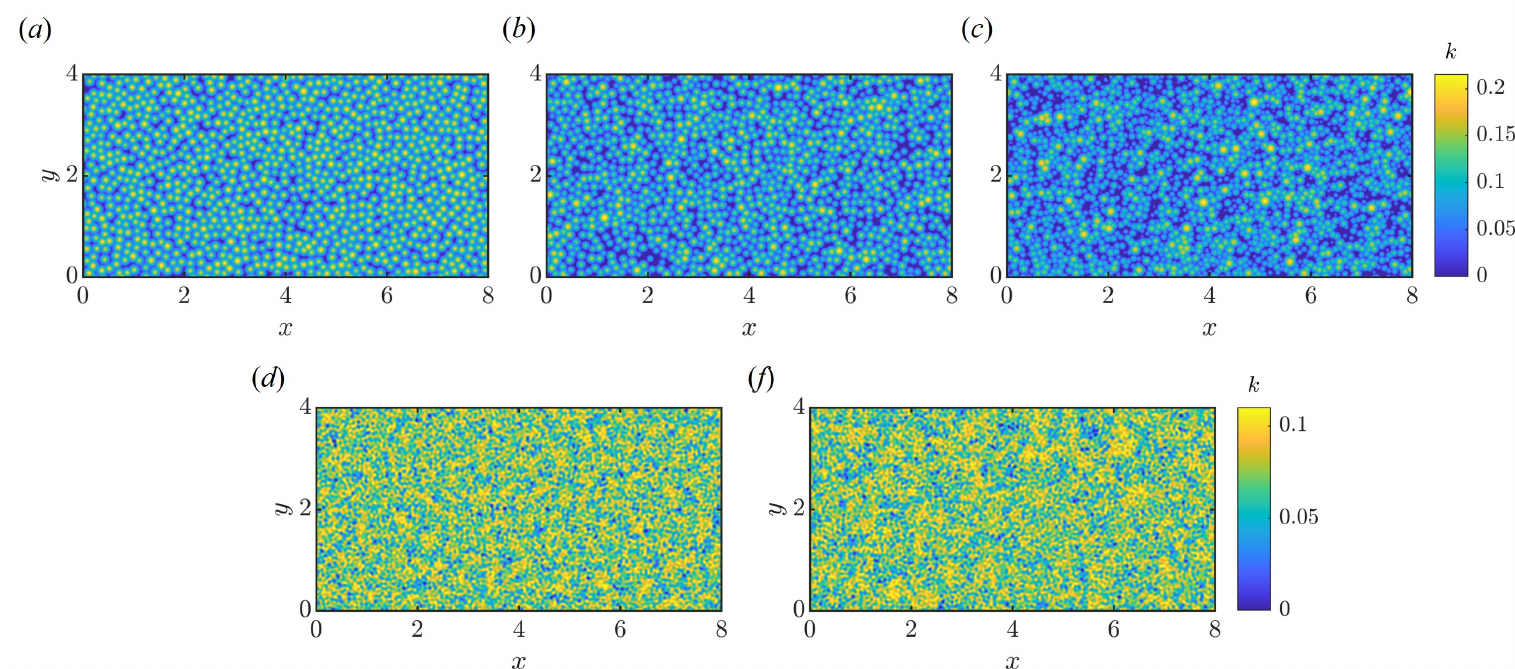}}
  \caption{Examples of surfaces with very similar statistics when considering the 7 roughness statistics in Table \ref{tab:DNS results}. The surfaces (a) A1588 (b) A3588 and (c) A7088 from \cite{Forooghi2017TowardCorrelation} belong to one group. The surfaces (d) N14 and (e) N18 from \cite{yang2022direct} belong to another group.}
\label{fig:surf_with_very_similar_statistics}
\end{figure}

A scenario might occur when this feature selection methodology when extended to a larger set of surfaces would exhaust all features. 
In this case, certain surfaces are indistinguishable from the available roughness statistics yet they have different $k_s$.
For example, figure \ref{fig:surf_with_very_similar_statistics} depicts two clusters of such surfaces from \cite{Forooghi2017TowardCorrelation} and \cite{yang2022direct} taken from a larger set of about 143 rough surfaces in the aforementioned roughness database.
The three rough surfaces in figure \ref{fig:surf_with_very_similar_statistics}(a), (b), and (c) from \cite{Forooghi2017TowardCorrelation} are referred to as A3588, A7088, and A1588. 
These are indistinguishable with the 7 roughness statistics in Table \ref{tab:DNS results}.
The $k_s/k_a$ magnitudes in surfaces A1588, A3588 and A7088 from \cite{Forooghi2017TowardCorrelation} stand around 11.1, 9.98, and 8.86 respectively, which are different, whereas the deviations in the aforementioned 7 statistics are less than 0.1 standard deviation (STD) for these surfaces.
The same applies to surfaces N14 and N18 from \cite{yang2022direct} whose $k_s/k_a$ magnitudes are close to 5.15 and 4.57 respectively.

\subsection{Building a rough-wall model}

Having identified the important roughness statistics required to build a rough-wall model, this section discusses the next step involved, which is developing rough-wall models themselves. 
This is quite straightforward, and there are a number of approaches.
Here, we present two such approaches: a multi-variate polynomial regression (MPR), which gives interpretable explicit expressions but has less descriptive power, and a feedforward neural network (FNN) based approach, which has great descriptive power but is a black box.

For the MPR model, the following steps were used to generate the polynomial terms: (i) minmax scaling and polynomial feature generation, (ii) determination of the degree and the polynomial features involved and (iii) linear regression of the polynomial features.
The regression follows an iterative bagging-based approach.  
At every iteration, the training dataset is randomly selected maintaining a test-train split of 0.7 and m random polynomial terms are selected to develop a regression fit. 
Ridge regression is employed here and the mean-squared test and training errors are computed. 
This step is repeated iteratively for different polynomials with the selected degree $d$ and $m$ number of polynomial features until we find a fit where the minimum in both training and test datasets occurs. 
These fits for the groups \cite{AghaeiJouybari2021Data-drivenFlows}, \cite{womack2022turbulent}, \cite{xu2021flow} and the current study are given by expressions \eqref{exp:Aga_Joey}, \eqref{exp:Womack}, \eqref{exp:Xu} and \eqref{exp:Current} respectively.
The reader is directed to \cite{kleinbaum2013applied} for further details of MPR.

\begin{multline}
         k_s/k_a = -0.53(k_{rms}/k_a)^3 -0.26Sk^3 +0.69 ES^3 +0.78(k_{rms}/k_a)^2\cdot Sk + 2.22Sk^2\cdot ES \\ + 2.45(k_{rms}/k_a)^2 \cdot ES +1.44(k_{rms}/k_a) \cdot Sk^2 -0.85Sk\cdot ES^2 +2.86(k_{rms}/k_a)\cdot Sk \cdot ES \\ -0.90(k_{rms}/k_a)^2 -1.51ES^2 + 0.5 Sk \cdot ES -0.44(k_{rms}/k_a)\cdot Sk +0.65ES -0.12k_{rms}/k_a
        \label{exp:Aga_Joey}
\end{multline}

\begin{equation}
     k_s/k_a = 0.95Sk^2 + 0.77ES\cdot Rl_x/k_a -2.63ES +1.79Sk -2.58k_{rms}/k_a
    \label{exp:Womack}
\end{equation}

\begin{equation}
     k_s/k_a = 0.88(k_{rms}/k_a)^2 + 0.13k_{rms}/k_a
    \label{exp:Xu}
\end{equation}

\begin{multline}
     k_s/k_a = 0.03{ES}^2  -1.19({Rl}_x/k_a)^2 + 1.00({Rl}_y/k_a)^2 -0.08({Rl}_x/k_a)\cdot({Rl}_y/k_a) \\ + 0.06ES + 1.71Rl_x/k_a -1.41Rl_y/k_a 
     \label{exp:Current}
\end{multline}

\begin{figure}
  \centerline{\includegraphics[width=0.9\linewidth]{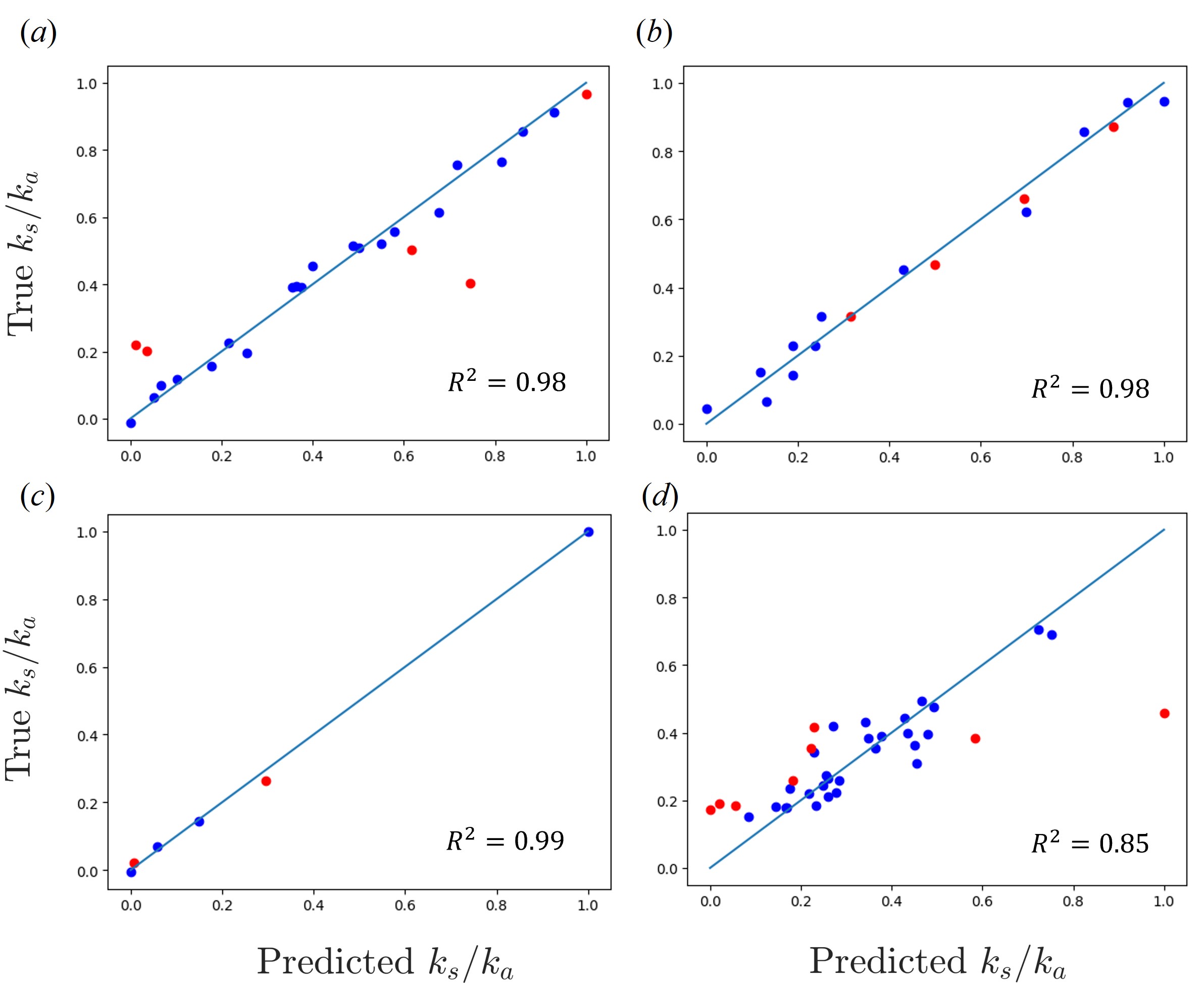}}
  \caption{MPR based regression for groups: (a) \cite{AghaeiJouybari2021Data-drivenFlows} (b) \cite{womack2022turbulent} (c) \cite{xu2021flow} and (d) Current study. Here, $R^2$ stands for coefficient of determination and is calculated on the training dataset. The red color stands for test data and the blue stands for training data. Here, $k_s/k_a$ is normalised using a MinMax scaler.}
\label{fig:fits}
\end{figure}

Figure \ref{fig:fits} shows the true vs predicted plots for the various groups. 
$R^2$ stands for coefficient of determination which is a measure of accuracy.
The $R^2$ shown in figure \ref{fig:fits} has been calculated from the training data which hold values 0.98, 0.98, 0.99 and 0.85 for the respective groups.
The corresponding $R^2$ based on the test data stand as 0.77, 0.99, 0.99 and 0.64. 
It is possible to choose polynomial forms of higher $R^2$ on the training data but we may risk overfitting.
Hence, these models are adopted maintaining a balance between generalization and accuracy.
It is further noted that the errors shown in $k_s$ predictions here would generate much smaller errors in drag prediction, as the skin-friction coefficient $C_f$ is proportional to [log$(k_s^+)]^2$.

\begin{figure}
  \centerline{\includegraphics[width=0.9\linewidth]{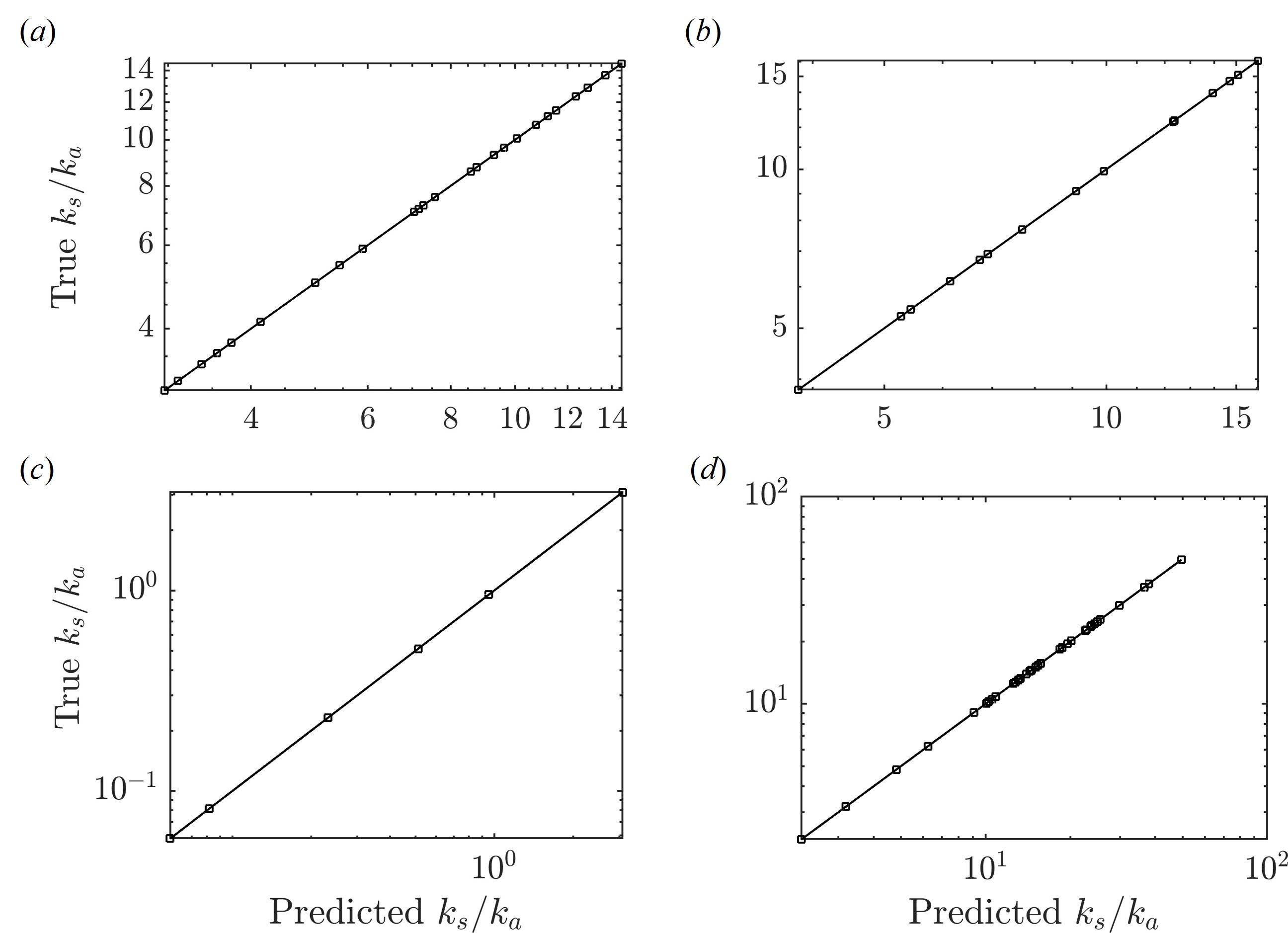}}
  \caption{FNN based regression for groups: (a) \cite{AghaeiJouybari2021Data-drivenFlows} (b) \cite{womack2022turbulent} (c) \cite{xu2021flow} and (d) Current study.}
\label{fig:NNfits}
\end{figure}

For the FNN model, a standard 3-layer network has been trained with 10 neurons in each layer. 
The size of the neural network is small but as we will see is sufficient.
Figure \ref{fig:NNfits} shows the parity plots for these networks. 
Here each FNN utilizes 100\% of its group for its training data. 
This is done to show the perfect fitting of the data in each case, which demonstrates the effectiveness of our feature selection strategy and the strong descriptive power of FNNs.

\section{Concluding remarks}\label{sec:5}

In conclusion, this study presents new data for rough surfaces in the relatively low skewness and high $k_{rms}/k_a$ space which have not been explored before.
The variations in equivalent roughness height, mean velocity profiles and Reynolds and dispersive stresses with changes in skewness ($Sk$), planar packing density ($\lambda_p$) and {roughness arrangement} have been reported. 
While the first two have been investigated in various studies, the latter effect of roughness arrangement is rarely considered.
In this work, this effect of roughness arrangement is not only considered but has been found to significantly affect the drag produced by rough surfaces.
This can be confirmed by the trends in the mean velocity profile.
In general, higher mean velocity profiles are observed for lower magnitudes of $Sk$ and $\lambda_p$ (except in AX surfaces where flow-sheltering is observed at higher $\lambda_p$).
For variations in roughness arrangement, the AXY and SXY orientations seem to produce more drag than the S and AY arrangements, followed by the AX arrangement.  
The valleys have been found to produce minimal, but non-zero contributions to drag, most of which we posit to stem from the form drag.
In the context of second-order flow statistics, the spatial flow inhomogeneity as indicated by dispersive stresses, is observed to be more in AX surfaces as compared to AY surfaces.
Secondary flow in the spanwise and wall-normal directions were also observed in the flow over these rough channels.

It is important to note that the effect of roughness element shape on the observed turbulence and drag characteristics has not been pursued in this study.
The choice of the cube-roughness geometry here is based on its simplicity to construct such surfaces in the $k_{rms}/k_a$ - $Sk$ parameter space and the ability to align the immersed solid boundaries with the cartesian mesh for most cases in this study.
While it is difficult to discuss the potential impact of roughness shape on the flow, the underlying assumption of the statistics-based rough-wall modelling approach is that the overall drag will be similar given the statistics are similar irrespective of the roughness element shape.

Since a majority of rough surfaces in the current study vary by rough-element arrangement, these are bound to have identical moments of roughness height. 
Two-point spatial correlation lengths have been proposed as input parameters to differentiate between these arrangements.
Additionally, we put forward the argument that based on the properties of the rough surfaces in the present dataset, finding a universal roughness correlation is an exacting task and that a "case-by-case" basis would do better.
We assert that roughness modelling can be viewed as determining input parameters that best distinguish the rough surfaces.
We demonstrate a methodology to identify these specific features for a given group of rough surfaces.
It is important to note that these distinguishing features vary based on the group being considered. 
While it might be useful to link a particular roughness feature as being important for a specific type of roughness, we would like to point out that such links may not necessarily find a physical basis, but rather a statistical one.

Having identified the important roughness features, we demonstrate both MPR and FNN-based approaches to build rough-wall models from the selected features. 
The selected feature can be found to be robust, given the goodness of fit observed in the models built from them.

\backsection[Funding]{This work is supported by NSF grant number 2231037 with Ronald Joslin as the technical monitor.}

\backsection[Declaration of interests]{The authors report no conflict of interest.}

\bibliographystyle{jfm}
\bibliography{references}

\end{document}